\documentclass[english]{revtex4-2}
\usepackage{color}
\usepackage[T1]{fontenc}
\usepackage[latin9]{inputenc}
\usepackage{geometry}
\usepackage{amsmath}
\usepackage{amsthm}
\usepackage{graphicx}
\PassOptionsToPackage{normalem}{ulem}
\usepackage{ulem}

\makeatletter

\def\tb{\textcolor{black}}

\def\bv{Brunt-V\"{a}is\"{a}l\"{a} }

\numberwithin{equation}{section}
\numberwithin{figure}{section}

\makeatother

\usepackage{babel}
\begin{document}
\title{Instability driven by settling and evaporation in a shear flow: a model for asperitas
clouds}
\author{S. Ravichandran}
\email{ravichandran@su.se}
\affiliation{Nordita, KTH Royal Institute of Technology and Stockholm University,
Stockholm, Sweden SE-10691}
\author{Rama Govindarajan}
\email{rama@icts.res.in}
\affiliation{International Centre for Theoretical Sciences, Bengaluru 560089, India.}
\begin{abstract}
We study, by direct numerical simulations in two and three dimensions, the instability caused by the settling and evaporation
of water droplets out of a cloudy layer saturated with vapour into
a dry sub-cloud ambient, under conditions where mammatus clouds were shown to form by 
\cite{Ravichandran2020}, but with the addition of background shear. We show that shear changes the type of cloud formation qualitatively, from mammatus-like to a newly identified cloud type called asperitas. Intermediate levels of shear are shown to be needed. Shear suppresses the growth of small-scale perturbations, giving rise to smooth, long-lasting structures, and smaller rates of mixing. Three-dimensionality is shown to make a qualitative difference, unlike in mammatus clouds. We also show that under non-cloud-like conditions, the instability can be very different. 
\end{abstract}
\maketitle

\section{Introduction \label{sec:Introduction}}

The distinctive shapes that atmospheric clouds take make them an artist's delight.
Among the most visually stunning of the various cloud types are mammatus clouds, which are pendulous blobs of cloud fluid suspended, as it were, from the sky. Mammatus clouds are typically seen underneath cumulonimbus anvils, although they also are known to form elsewhere.
Several explanations for their formation, and for their characteristic appearance, have been proposed, including
layer subsidence, radiative cooling and the evaporation of ice particles
and/or water droplets settling out of the cloud. See \citep{Shultz2006}
for a comprehensive review, and \citep{Kanak2006,Kanak2008} for numerical
simulations of mammatus clouds suggesting that the cooling of the
sub-cloud air plays a crucial role in the formation of mammatus clouds.
It has also been suggested that mammatus clouds are simply the descending
part of the circulation created due to the radiative temperature difference
between the lower boundary of the cloudy layer and the sub-cloud dry
air \citep{Garrett2010}. Ravichandran {\em et al.} \citep[][hereafter RMG20]{Ravichandran2020} proposed that mammatus clouds form due to the settling of liquid droplets into the dry air layers beneath the cloud, resulting in evaporative cooling of this layer, and setting into motion a Rayleigh-Taylor-type instability. They studied the role
of droplet size and liquid water content on the lobe sizes that can
result using linear stability analysis and direct numerical simulations,
finding that lobe-like instabilities result only for sufficiently
large droplet sizes as well as large liquid water content. It has been pointed out, however, that none of
the proposed mechanisms explains every observation of mammatus clouds,
and therefore that more than one mechanism can cause lobe-like clouds
that may be called mammatus \citep{Shultz2006}. 

\begin{figure}
    \centering
    \includegraphics[width=0.46\columnwidth]{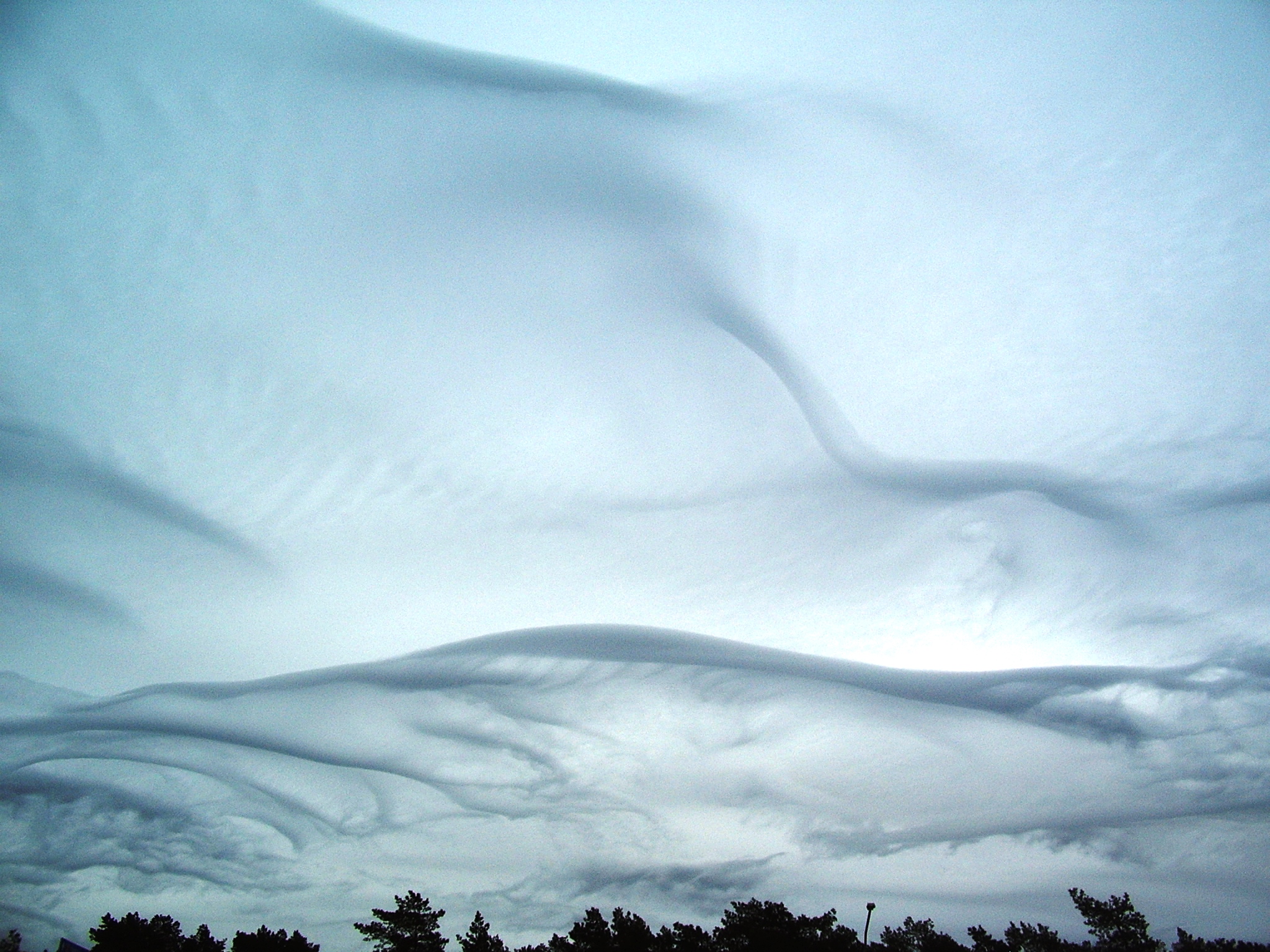}
    \includegraphics[width=0.51\columnwidth]{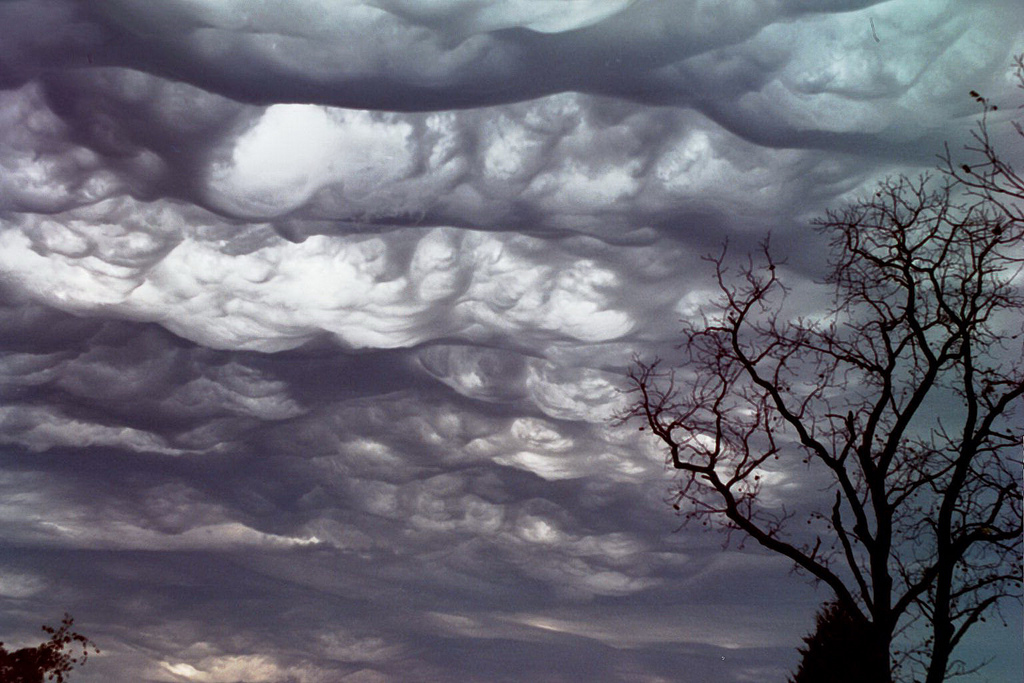}
    \caption{\label{fig:asperitas_web_pics}
    Left: Asperitas clouds over Tallinn, Estonia, on April 9th, 2009. 
    By Ave Maria M{\~o}istlik, Own work, CC BY-SA 3.0 \cite{AveMaria}.
    Right: Asperitas clouds over Pocahontas, Missouri, on August 9th, 2008. 
    By Agathaman, Own work, CC BY-SA 3.0 Unported \cite{Agathaman}.
    }
\end{figure}

In the year 2017, for the first time in nearly six decades, the World Meteorological Organisation (WMO) designated a new cloud type called ``Asperitas'' \footnote{https://cloudatlas.wmo.int/en/clouds-supplementary-features-asperitas.html}, from the Latin
for `severity', described to be ``well-defined, wave-like structures in the underside of the cloud; more chaotic and with less horizontal organization than the variety undulatus. Asperitas is characterized by localized waves in the cloud base, either smooth or dappled with smaller features, sometimes descending into sharp points, as if viewing a roughened sea surface from below.'' The wave-like nature and the descent into sharp points suggest that their creation process could be related to that of mammatus clouds. They however look wavy rather than blob-like and the lobes are inclined to the vertical. Two typical examples, pictures available in the public domain, are reproduced in figure \ref{fig:asperitas_web_pics}. Several more may be found on the website of the Cloud Appreciation Society (whose efforts were responsible for the designation of Asperitas as a separate cloud type) at https://cloudappreciationsociety.org/ A stunning timelapse of asperitas evolution may be found here: https://www.youtube.com/watch?v=Jz7BgxrVmiQ . 

The only paper we can find on asperitas cloud is by
\citep{GilesHarrison2017} (hereafter GH17), based on a dissertation
by one of the paper's authors. GH17 find that in the few complete
observations of asperitas clouds available, the Richardson number
$Ri=N^{2}/S^{2}>0.25$, where $N$ is the \bv frequency
and $S$ the shear rate, is too large for the Kelvin-Helmholtz instability to be active. They mainly study observations, and also perform some idealised large eddy simulations using radiosonde profiles associated with the observations of asperitas as initial conditions. They propose that layers of stratification within the cloud can give rise to ducted gravity waves under certain conditions, and that this could form asperitas clouds, aided by differential shear in the layers. We propose a different mechanism
for the formation of asperitas clouds based on turbulence and cloud microphysics: by the settling and evaporation of water droplets in a background shear. The thermodynamics of phase change is an important ingredient in our mechanism. Since this cloud type is only recently identified, further work will be needed to critically examine the applicability of each of these two mechanisms, and perhaps others. We begin by discussing settling-driven instabilities in various contexts without phase change.

Fluid dynamical systems with multiple scalar components where one
of the scalar components has a finite settling velocity (typically
because it is a suspended phase) occur frequently in geophysics. For
instance, in estuaries where silt-laden river water flows into the
sea, the rate of mixing of the fresh water into the saline water is
mediated by the settling driven instability at the interface (see,
e.g. \citep{Burns2012,Burns2014, Yu2013, Yu2014}). Settling driven instabilities (with the additional complexity
of coagulation and a changing diffusivity of the suspended phase)
are also responsible for layer formation in volcanic ash mushroom
clouds \citep{Carazzo2012,Carazzo2013}, and the significant increase
in the lifetime of such ash clouds that results.

As we shall see, settling-driven instabilities are broadly related to double-diffusive (DD) instabilities, where the system in the base state is nominally stably stratified, but instabilities occur because of differential diffusion of different scalars in the flow. The difference is that in the typical DD system, none of the scalar components undergoes settling. Both DD and settling driven instabilities have been studied in the presence of a horizontal shear flow \citep{Radko2015,Konopliv2018,Garaud2019,Sichani2020}. In general, due to the imposed homogeneity in the direction of flow, shear stabilizes the fingering instabilities that occur at the interface, while modes in the direction perpendicular to the shear are unaffected. In double-diffusive systems, this leads to the formation of so called `salt sheets' which have been observed experimentally and in simulations, reducing the vertical fluxes of heat and salinity (see, e.g. \citep{Radko2015}).

Here, we study the influence of shear on the lobe-like instabilities
that arise due to the settling and evaporation of water droplets. Our system has two non-settling scalar components, namely temperature and water vapour, in the ambient fluid (air) and liquid droplets which settle under gravity, and undergo evaporation as they do so. Without a base shear, mammatus clouds form under a range of physical parameters, and in a certain range of shear, we propose that our mechanism supports the formation of asperitas clouds.

We employ the formulation used in RMG20 and coarse-grain
the liquid water droplets into a scalar field with a finite settling
velocity which can vary in space and time. We perform direct numerical
simulations (DNS) of the governing equations in two and three dimensions,
with the interface between cloudy and clear air initially perturbed
sinusoidally with a given single wavelength, and an externally imposed
shear flow maintained by upper and lower boundaries that move in opposite
directions, and present results from these simulations. 

The rest of the paper is organised as follows. In section \ref{sec:Setup},
we describe the geometry of the problem and outline the derivation
of the governing equations. We then nondimensionalise the equations
and list the governing nondimensional parameters. In section \ref{sec:Results},
we present and discuss results from DNS in two and three dimensions (2D and 3D), showing that
the extra spatial dimension changes the dynamics qualitatively when
shear is nonzero. We then conclude in \ref{sec:Conclusion} with some
thoughts on future work.

\section{Problem Setup and Numerical Simulation \label{sec:Setup}}

\begin{figure}
\begin{centering}
\includegraphics[width=0.6\paperwidth]{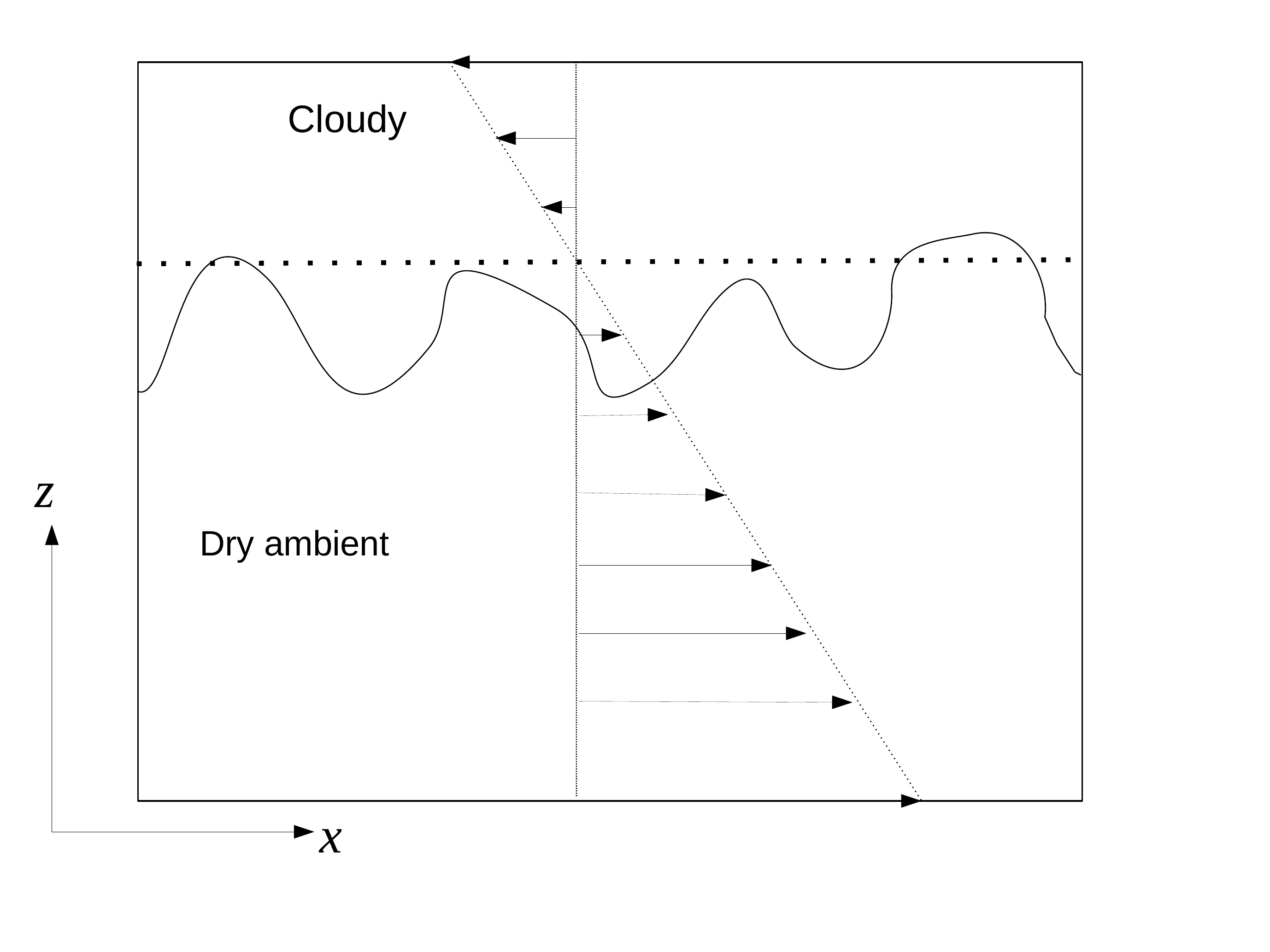}
\par\end{centering}
\caption{\label{fig:schematic} The geometry of the problem, shown here in two-dimensions. The background shear is linear, with zero velocity at the location $z=z_0$ (the dotted line) and the velocity for $z<z_{0}$ chosen to be positive in the $x$-direction. At the initial time, the region $z>z_{0}$ contains water vapour at saturation and liquid water droplets, whereas the domain below $z_0$ is dry. The interface is initially perturbed (solid wavy line) as discussed in the text. Gravity points in
the $-z$ direction.}
\end{figure}
We devise the simplest initial scenario for the formation of asperitas clouds, shown schematically in figure \ref{fig:schematic}. The geometry is a rectangular or cuboidal volume of dimensions $L_{x},\left(L_{y}\right),L_{z}$. We show $L_y$ in brackets since it appears in our three-dimensional simulations, whereas our two-dimensional simulations are in the $x-z$ plane. The domain consists of a layer of cloud fluid placed above a layer of dry air. The cloud fluid is a mixture of dry air, water vapour and liquid water droplets. 
Initially, the mean location of the interface between cloudy and clear air is $z_{0}$. The entire system is under constant shear $\tilde{S}$, with the initial horizontal component of velocity $u$ varying linearly with the vertical coordinate as
\begin{equation}
u(z,t=0)={\tilde{S} }\left(z_{0}-z\right).\label{eq:shear_0}
\end{equation}
On the lower and upper boundaries, at $z=0$ and $z=L_{z}$ respectively, we apply horizontal velocity boundary conditions matching the shear profile, along with setting the spanwise and vertical velocities $v$ and $w$ to be zero: 
\begin{align}
u\left(z=0\right) & =\tilde{S} z_{0}\nonumber \\
u\left(z=L_{z}\right) & =-\tilde{S}\left(L_{z}-z_{0}\right)\nonumber \\
v\left(z=0\right)=v\left(z=L_{z}\right) = w\left(z=0\right)=w\left(z=L_{z}\right) & =0.\label{eq:vel_BCs_0}
\end{align}
The horizontal directions are periodic. The system is initially isothermal, with the temperature $T=T_b$ everywhere, and the initial vapour concentration in the cloud layer is set at the saturation value for this temperature. The interface at $z=z_{0}$ is initially perturbed sinusoidally with an amplitude $h/2$ \tb{and wavelengths $\lambda_x$ and $\lambda_y$ in the $x$ and $y$ directions respectively (in 2D, the last factor in \eqref{eq:interface_0} is set identically to unity)}:
\begin{equation}
z_{i}=z_{0}+\frac{h}{2}\text{sin}\left(\frac{2\pi x}{\lambda_x}\right)\tb{\text{sin}\left(\frac{2\pi y}{\lambda_y}\right)},\label{eq:interface_0}
\end{equation}
with the region $z>z_{i}$ containing saturated vapour and liquid water droplets, and $z<z_{i}$ consisting of dry air.
The water droplets of initial radius $a_{0}$ and number density $n$ in the cloudy part of the domain, represented here in coarse-grained form by a liquid field. 
The quantities of vapour and liquid water are given in term of their normalised mixing ratios (mass of vapour/liquid per unit mass of dry air):
\begin{equation}
    r_{v,l} =  \frac{\tilde{r}_{v,l}}{r_s^0}  = \frac{\rho_{v,l}}{\rho_d r_s^0}, \label{eq:defn_rv_rl}
\end{equation}
where $\rho_{v,l}$ are the densities of vapour and liquid, $\rho_d$ is the density of dry air. We have normalised the 
mixing ratios $\tilde{r}_{v,l}$ using $r_s^0 = \rho_s^0 / \rho_d$, the saturation vapour mixing ratio at 
the base temperature $T_b$ (see RMG20 for a more detailed discussion). The deviation from $T_b$ everywhere in the flow is represented by a nondimensional temperature
\begin{equation}
    \theta=\frac{T-T_b}{\Delta T}
\end{equation}
We thus have, at $t=0$,
\begin{align}
r_{v,0}\left(z\right) & =\begin{cases}
1 & z\geq z_{i}\\
0 & z<z_{i}
\end{cases},\label{eq:vap_0}
\end{align}
\begin{equation}
r_{l,0}\left(z\right)=\begin{cases}
r_{l}^{0}+\text{noise} & z\geq z_{i}\\
0 & z<z_{i}
\end{cases},\label{eq:liq_0}
\end{equation}
and
\begin{equation}
    \theta_0=0. \label{eq:theta_0}
\end{equation}
The subscript $0$ indicates a quantity at time $t=0$.

We assume that the water droplets are small enough ($\mathcal{O} (10) \mu$m) to remain spherical and obey the Stokes drag law. They therefore settle with a finite instantaneous velocity 
\begin{equation}
    v_{p}\left(a\right) = \frac{2g\rho_w}{9\nu\rho_d} a^{2}, \label{eq:vp} 
\end{equation}
where $a$ is the instantaneous droplet radius and $g$ is the acceleration due to gravity.
We note that droplet sizes observed in real asperitas clouds (which, from the very limited observations available \cite{asperitas_conditions}, seem to form under similar atmospheric conditions as mammatus)  are larger, and the Stokes law, equation \ref{eq:vp}, will need higher order corrections. \tb{Large droplets are also subject to inertial effects, which cause the droplets to deviate from fluid streamlines, with the droplet velocities $\boldsymbol{v}$ given by}
\[
\tb{\boldsymbol{v} = \boldsymbol{u} - \frac{v_p}{g}\frac{D\boldsymbol{u}}{Dt}},
\]
\tb{where $D/Dt$ is the material derivative following fluid streamlines and we have expanded the Maxey-Riley equations \cite{Maxey1983} for a heavy particle to first order in the particle timescale $\tau_p = v_p / g$ (which is small relative to the flow timescales). Here, following refs.  \cite{Yu2013,Yu2014,Burns2014}, we ignore these effects, while retaining the settling velocity of the droplets. }

As in RMG20, we assume that $n$ is constant everywhere i.e., as droplets evaporate, their number density remains constant even as their radii shrink. The instantaneous radius of the droplets is prescribed by its representative value $a = [3 \rho_l/(4 \pi n)]^{1/3}$ inside each grid cell. We note that droplet radii are far smaller than the Kolmogorov scale, and in a typical cloud, the number density is $O(1)$ in a cube whose side is the Kolmogorov scale. The same number density is prescribed nominally in the initially dry region in the lower half of the domain, although wherever $r_l=0$, it does not affect the dynamics. Droplets evaporate on a timescale (see also equation \ref{eq:condensation} below)
\begin{equation}
    \tilde{\tau}_{s} = \frac{C r_s^0 \rho_d}{4\pi n a} = \frac{C \rho_w}{3} \frac{a^2}{r_l} \label{eq:taus_dim}
\end{equation}
 where $r_l$ is the normalised liquid mixing ratio (equation \ref{eq:defn_rv_rl}) and $C$ is a dimensional constant whose value is about $10^7 sec \ m  \ kg^{-1}$ in a cloud. The velocity scaling emerges out of considering the droplets as spheres in the Stokesian limit, whereas the evaporation time scale arises from assuming purely diffusive vapour flux from isolated spherical droplets \cite[e.g][]{Pruppacher2010}. As they settle out of the cloud into the dry ambient, droplets evaporate, cooling and humidifying the subsaturated air underneath the cloudy layer. 

The governing equations are the Boussinesq Navier-Stokes equations with advection-diffusion equations for $\theta$, $r_v$ and $r_l$.
To nondimensionalise the equations, we must define a length scale. The total height of the cloud layer could be a possibility, but this 
height does not directly influence the physics under consideration. We therefore choose the end-to-end initial perturbation amplitude $h$, 
as the length scale. The initial settling velocity $v_{p,0}$ provides a natural velocity scale, giving
a timescale $h/v_{p,0}$. The nondimensional governing equations are 
\begin{align}
\nabla\cdot\boldsymbol{u} & =0,\label{eq:continuity}\\
\frac{D\boldsymbol{u}}{Dt} & =-\nabla p+\frac{1}{Re}\nabla^{2}\boldsymbol{u}+\frac{\boldsymbol{e}_{z}}{Fr^{2}}\left[\theta+r^{0}\left(\chi r_{v}-r_{l}\right)\right],\label{eq:momentum}\\
\frac{D\theta}{Dt} & =\frac{1}{Re}\nabla^{2}\tb{\theta}+L_{1}C_{d},\label{eq:temperature}\\
\frac{Dr_{v}}{Dt} & =\frac{1}{Re}\nabla^{2}\tb{r_v}+C_{d},\label{eq:vapour}\\
\frac{Dr_{l}}{Dt} & =C_{d}+ \frac{v_p}{v_{p,0}}\frac{\partial r_{l}}{\partial z},\label{eq:liquid}
\end{align}
where \tb{$r^0 = r_s^0 T_b / \Delta T$ is a ratio of buoyancy contributions from water substance and temperature, $1+\chi=28.9/18\approx1.6$ is the ratio of molecular masses of air and water vapour,} and $C_{d}$ is the nondimensional rate of condensation
\begin{equation}
C_{d}=\frac{dr_{l}}{dt}=\frac{\mathcal{H}(r_v,r_l,r_s)}{\tau_{s}}\left(\frac{r_{v}}{r_{s}}-1\right)\label{eq:condensation}
\end{equation}
where $\tau_{s}=\tilde{\tau}_s v_{p,0}/a_0$  denotes the instantaneous nondimensional evaporation timescale, \tb{and $\mathcal{H}(r_v,r_l,r_s)$ is a Heaviside function} indicating if evaporation or condensation occur, given by
\begin{equation}
\tb{\mathcal{H}(r_v,r_l,r_s) = }\begin{cases}
1 & r_v \geq r_s \mathrm{\ \  or\ \  } r_l > 0\\
0 & \mathrm{otherwise}
\end{cases}
\end{equation}
and the (normalised) saturation vapour mixing ratio $r_{s}$ is given by the Clausius-Clapeyron law
\begin{equation}
r_{s}=\text{exp}\left(L_{2}\theta\right).\label{eq:Clausius}
\end{equation} 
Note that the Reynolds number appears in the equations for temperature and water vapour as well. This is because we have chosen both the Prandtl and Schmidt numbers to be $1$, which is realistic for these scalars. The Schmidt number for the liquid  is taken to be infinite. \tb{Similar equations have been proposed to study the formation of precipitating convection \citep{Hernandez2013} where vapour condenses to form liquid water droplets that can rain. Other simplified models include the model of ref. \cite{Pauluis2010} which assumes a piecewise-linear equation of state, and the `rainy B\'enard' model of ref. \cite{Vallis2019} where the full nonlinear equation of state is used, but the settling velocity of droplets is neglected.}
\tb{We also note that the initial conditions given by equations (\ref{eq:vap_0}--\ref{eq:theta_0}) are only steady-state solutions of the governing equations (\ref{eq:continuity}--\ref{eq:liquid}) if both the diffusivity of vapour and the settling velocity of liquid water are switched off. }

Following RMG20, we allow the settling velocity and the evaporation timescale to change as the liquid field evolves, 
by maintaining  the number density of droplets $n$ to be constant during a simulation, such that $\tau_{s} = \tau_{s,0} \left[r_{l}/r_{l}^0\right]^{-1/3}$
and $v_{p} = v_{p,0} \left[r_{l}/r_{l}^0\right]^{2/3}$, where $\tau_{s,0}$ and $v_{p,0}$ are the values for $a=a_0$ and $r_l=r_l^0$.
Choosing a base temperature $T_b$ and a temperature scale $\Delta T$,
and assuming the two phases involved are water and air (see below),
fixes the thermodynamic quantities 
\begin{equation}
L_{1} \equiv \frac{L_{v}r_{s}^{0}}{C_{p}\Delta T} \quad
{\rm and } \quad
L_{2} \equiv \frac{L_{v}\Delta T}{R_{v}T_b^{2}}.\label{eq:def_L1_L2}
\end{equation}
Here, $C_p$ is the specific heat capacity of air, $L_v$ is the enthalpy of vaporisation of water, and $R_v$ is the gas constant for water vapour.
The other nondimensional parameters controlling the dynamics are the
Froude number $Fr$, the Reynolds number $Re$ and the nondimensional shear rate $S$, given respectively by
\begin{equation}
Fr^{-2} \equiv \frac{g\Delta T/T_bh}{v_{p,0}^{2}},
\quad
Re \equiv \frac{v_{p,0}h}{\nu}, \quad {\rm and} \quad
S\equiv \frac{\tilde{S}h}{v_{p,0}},\label{eq:Fr2_Re_S}
\end{equation}
where $\nu$ is the kinematic viscosity of air. The Reynolds number in this study is held fixed at $Re=1000$. \tb{The nondimensional shear-rate $S$ defined here is the inverse of the nondimensional settling velocity $S_v$ defined, e.g., in ref. \cite{Devenish2012}. Typical values of the settling velocity for droplets in mature cumulonimbus anvils are $O(1-5)m/s$, and observations of asperitas clouds \citep{GilesHarrison2017} suggest $\tilde{S}=O(0.01-0.02)s^{-1}$. Taking $h=O(10)m$, we have $S=O(0.1)$, consistent with the values used here. Since the nondimensional settling velocity is larger than unity, the effects due to the settling and evaporation occur on a faster timescale than that associated with the shear. This suggests that the instability due to shear is a secondary instability that modifies the mammatus instability of RMG20, and that parallels may be drawn to the influence of shear on Rayleigh-Taylor or Holmboe instabilities \cite{Carpenter2011}.}

Using equations \ref{eq:vp} and \ref{eq:taus_dim} at the initial time, we have
\begin{equation}
Fr^{-2}\tau_{s,0} = \frac{g\Delta T/T_bh}{v_{p,0}^{2}} \frac{C \rho_w a_0^2}{3 r_l^0}\frac{v_{p,0}}{h} = \frac{3 \nu \rho_d C \Delta T/T_b}{2   \rho_w r_l^0}. \label{eq:clouds_Fr2_taus}
\end{equation}
Except for the temperature scale $\Delta T$, the base temperature $T_b$, and the initial liquid water content $r_l^0$ in the cloud, the quantities on the right hand side of equation \ref{eq:clouds_Fr2_taus} are properties of the air-water system. Our simulations are performed for $r_l^0=0.3$, $\Delta T = 1 K$ and $T_b=273K$ and thus the product $Fr^{-2} \tau_{s,0}$ must have a fixed value, about $2$ in our case, to be consistent with air-water properties in a cloud.
\begin{figure}
\begin{centering}
\includegraphics[width=0.3\paperwidth]{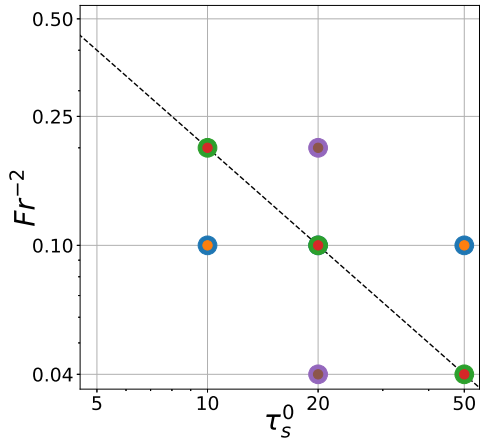}
\par\end{centering}
\caption{\label{fig:cases} A section of the parameter space explored in this
work. The dashed line is the curve $Fr^{-2}=2\tau_{s,0}^{-1}$, corresponding
to the behaviour of water droplets in clouds, and the green-red markers
are the specific combinations used. The blue-red markers are more
general combinations. For each combination of $\left(Fr^{-2},\tau_{s,0}\right)$ shown by the symbols,
we perform simulations for several combinations of the shear rate $S$ and the initial perturbation wavelength $\lambda_x$.}
\end{figure}
 
Our aim is to study how the instability that occurs due to this settling
and evaporation of the liquid water droplets, giving lobe-like
structures for large $v_{p,0}$ and $r_{l,0}$ \citep{Ravichandran2020},
is modified by the presence of shear. In particular, we ask whether shear can qualitatively change the very type of cloud type mammatus to asperitas. The requirement of significant shear could then explain the relative rarity of observations
of asperitas clouds and therefore their very recent designation as a separate
cloud type (as opposed to mammatus clouds which have been documented
for over a century). We report results from two- and three-dimensional
numerical simulations below.

The simulations are performed using the finite volume solver \emph{Megha-5},
with second-order central differences in space, and a second order Adams-Bashforth timestepping
scheme. \tb{For the liquid water equation \ref{eq:liquid}, where the diffusion term is absent, we implement the Kurganov-Tadmor scheme \citep{Kurganov2000} which enables non-diffusive advection, allowing for sharp gradients without incurring Gibbs oscillations. The numerical method was also described previously in RMG20.}  We fix $Re=1000$. In 2D, the domain is of size $80\times 80$ with
a grid of $1024\times1024$ points, while in 3D the domain is of size $40\times40\times80$
with $256\times256\times512$ grid points. The timestep is $dt\leq0.001$,
such that the Courant-Friedrichs-Lewy number $CFL \leq 0.3$ even for the largest shear rates. We have checked that the results do not change qualitatively upon doubling the number of grid points.

\section{Results and Discussion \label{sec:Results}}

\subsection{\tb{Linear Stability Analysis}\label{sec:linstab}}

\begin{figure}
    \centering
    \includegraphics[width=0.5\columnwidth]{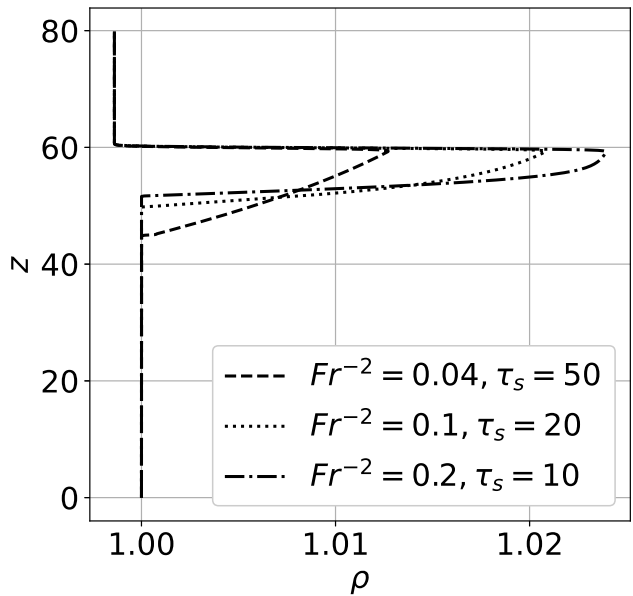}
    \caption{\tb{The 1D density profiles for the three cloud-like cases considered (see figure \ref{fig:cases}). As in RMG20, smaller droplet sizes lead to narrower overhangs with a larger maximum density. \label{fig:rho_1d}}}
\end{figure}

\tb{The governing equations \ref{eq:continuity}--\ref{eq:liquid} have a discontinuity at saturation. In order to perform linear stability analysis, therefore, we follow the same procedure as in RMG20. We run one-dimensional nonlinear simulations in the vertical coordinate and time  for the settling and evaporation of liquid water into a dry sub-cloud layer, and use the resulting density profile for our stability analysis. These profiles are slowly varying in time, and for the purposes of the stability analysis, can be assumed to be quasi-steady. More details are available in RMG20. These density profiles, shown in figure \ref{fig:rho_1d}, consist of `overhangs' of heavier fluid sandwiched between layers of lighter fluid at the cloudy-dry interface, with larger magnitudes and smaller widths for larger $Fr^{-2}$ and smaller $\tau_s$ (i.e. smaller droplet sizes). }

\tb{We examine the stability of the resulting `base state' by linearising the nondimensional governing equations for the vorticity }
\[
\tb{q=\frac{\partial u}{\partial z}-\frac{\partial w}{\partial x}}
\]
\tb{and the density $\rho$, which are}
\begin{align}
\tb{\frac{\partial q}{\partial t}+U\frac{\partial q}{\partial x}} & =\tb{\frac{1}{Re}\nabla^{2}q+\frac{1}{Fr^{2}}\frac{\partial\rho}{\partial x}\label{eq:q}}\\
\tb{\frac{\partial\rho}{\partial t}+U\frac{\partial\rho}{\partial x}} & = \tb{\frac{1}{Re}\nabla^{2}q-w\left(D\bar{\rho}\right),}\label{eq:rho}
\end{align}
\tb{where $D\equiv d/dz$, and $U\equiv U\left(z\right)=S(z_0-z)$ is the background shear velocity, while $\bar{\rho}\left(z\right)$ is the background density obtained from one-dimensional nonlinear simulations of the governing equations. }

\begin{figure}
    \centering
    \includegraphics[width=0.45\columnwidth]{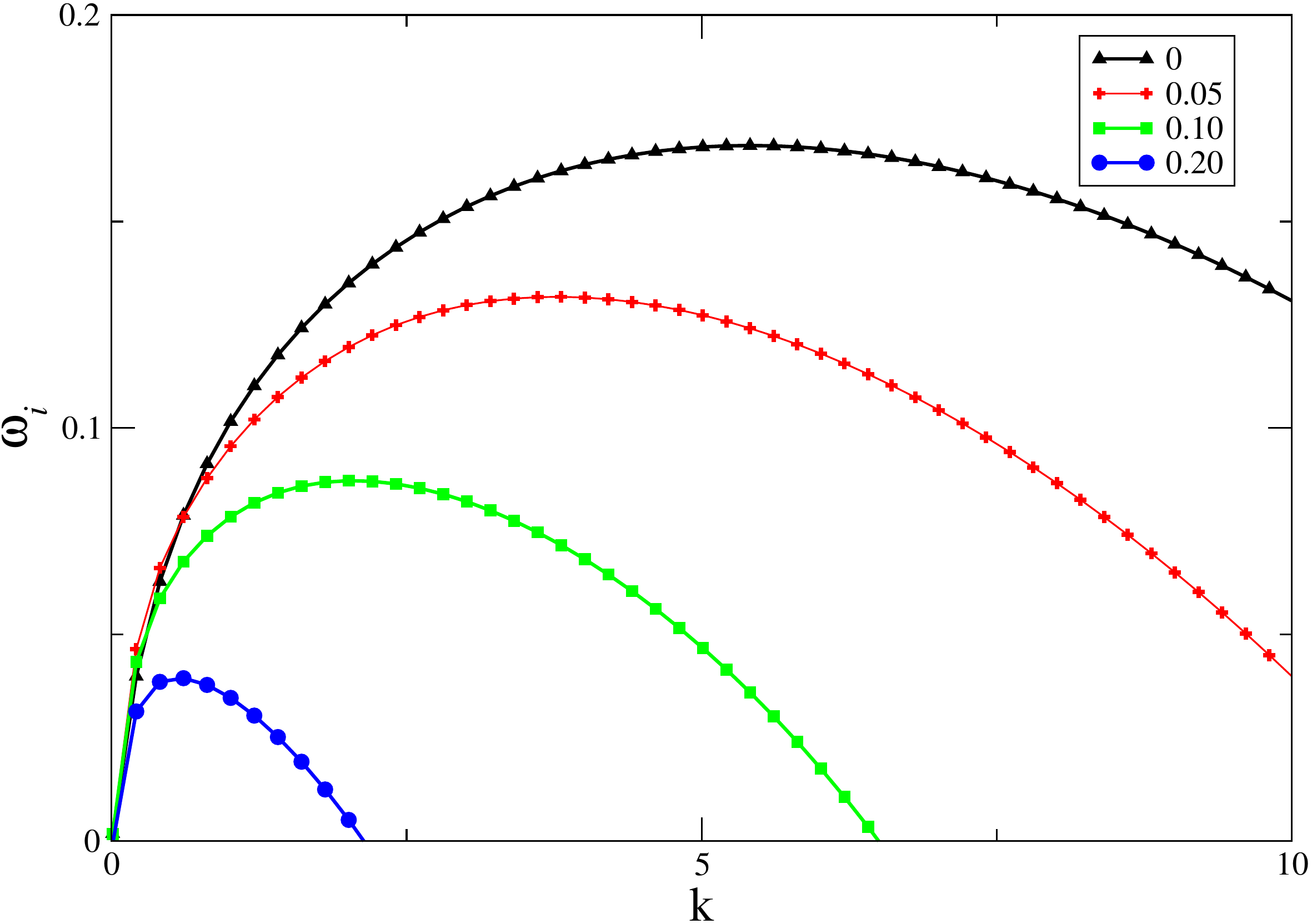}
    \includegraphics[width=0.45\columnwidth]{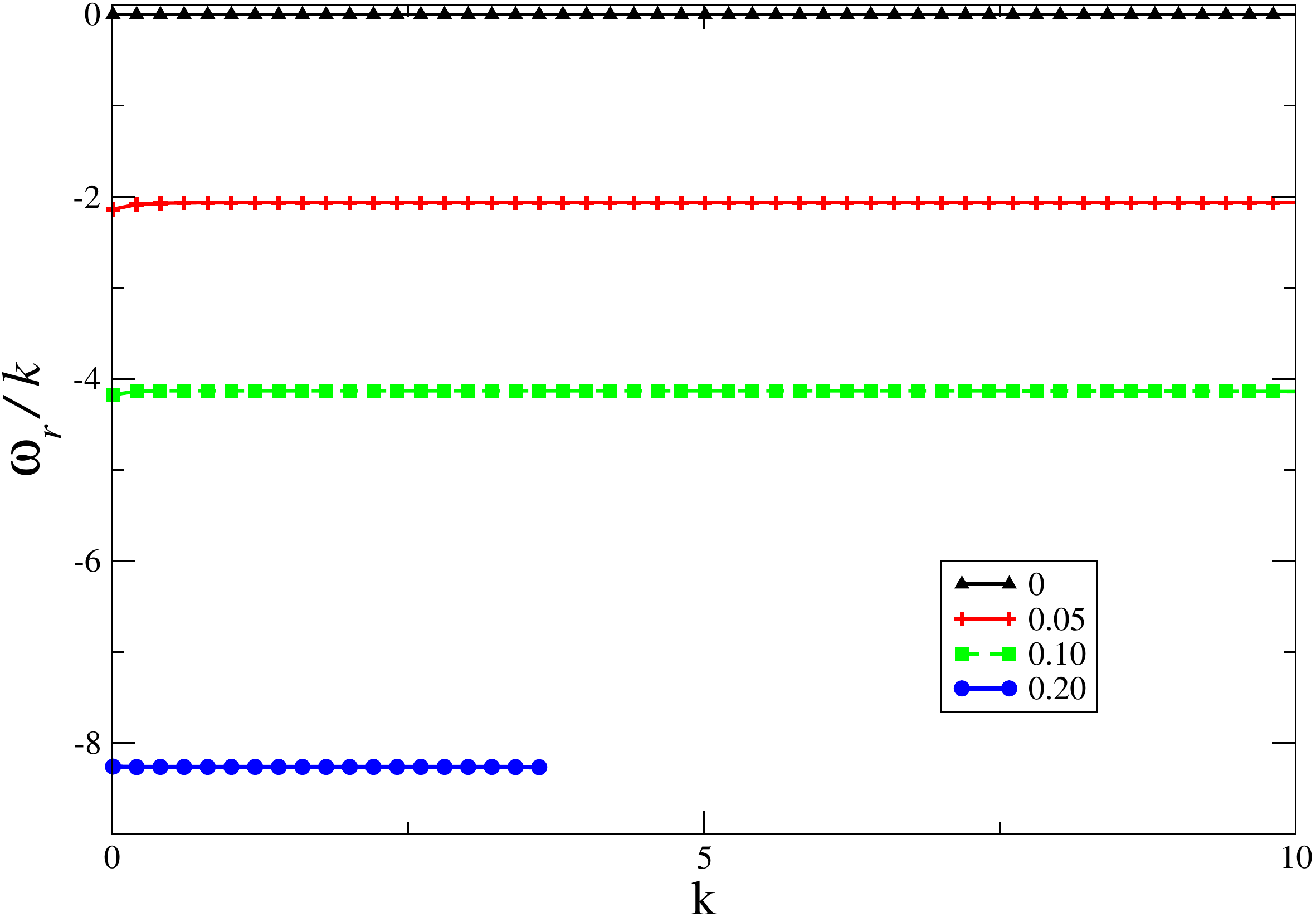}
    \caption{\tb{(a) The growth rate $\omega_i$ and (b) the phase velocity $\omega_r/k$ as a function of the horizontal perturbation wavenumber $k$ and the shear rate $S$, for density profiles obtained with parameters $Fr^{-2}=0.2$, $\tau_s = 10$. Results for other combinations of parameters show the same qualitative behaviour. \label{fig:lin_stab}}}
\end{figure}

\tb{Expanding $q$ and $\rho$ in normal modes}
\begin{equation}
\tb{\left(q,\rho\right)=\left(\hat{q},\hat{\rho}\right)\text{exp}\left[i\left(kx-\omega t\right)\right],\label{eq:normal_modes}}
\end{equation}
\tb{and noting that incompressibility requires }
\begin{equation}
\tb{ik\hat{u}+D\hat{w}=0,\label{eq:incompressibility}}
\end{equation}
\tb{gives}
\begin{equation}
\tb{\hat{q}=\frac{i}{k}\left(D^{2}-k^{2}\mathcal{I}\right)\hat{w}.\label{eq:q_from_w}}
\end{equation}
\tb{Equations \ref{eq:q} and \ref{eq:rho} in normal form, written for
$\hat{w}$ and $\hat{\rho}$ become}
\begin{align}
\tb{\frac{\partial}{\partial t}\left(D^{2}-k^{2}\mathcal{I}\right)\hat{w}+ikU\left(D^{2}-k^{2}\mathcal{I}\right)\hat{w}} & = \tb{\frac{1}{Re}\left(D^{2}-k^{2}\mathcal{I}\right)^{2}\hat{w}+\frac{ik}{Fr^{2}}\hat{\rho}}\label{eq:w_hat}\\
\tb{\frac{\partial}{\partial t}\hat{\rho}+ikU\hat{\rho} } & = \tb{\frac{1}{Re}\left(D^{2}-k^{2}\mathcal{I}\right)^{2}\hat{\rho}-\hat{w}\left(D\bar{\rho}\right).}\label{eq:rho_hat}
\end{align}
\tb{We see that the only imaginary terms are due to the background shear $U\left(z\right)$, as expected. Equations \ref{eq:w_hat} and \ref{eq:rho_hat} are solved for the complex growth rate $\omega$ using the Chebyshev decomposition employed in RMG20. The growth rate and the dispersion relation are plotted in figure \ref{fig:lin_stab}. Figure \ref{fig:lin_stab}(a) shows that the maximum growth rate occurs for smaller wavenumbers with increasing shear; thus, shear damps out small-scale instabilities. Figure \ref{fig:lin_stab}(b) shows that the phase speed scales with the shear rate $S$. We note that the flow velocity and the phase velocity coincide (i.e. $U(z)=\omega_r/k$) at $z\approx z|_{\rho_\textrm{max}}$. It is this interaction at the critical layer that leads to the instability. }
\tb{We note that while the linear stability results are qualitatively consistent with the results from nonlinear simulations to be presented below, the caveats described in RMG20 apply here too, so we do not make quantitative comparisons. In particular the base flow itself evolves over time, and we have used a density profile at an arbitrary time to define a  the linear stability analysis.}

\subsection{\tb{Results from 2D simulations}} \label{sec:2D}

RMG20 find that, in the absence of shear, mammatus lobe formation
can be predicted from 2D simulations, and nothing changes qualitatively between these and 3D simulations. We therefore begin our investigations
with 2D simulations. The application of shear sets a preferred
horizontal direction and, as discussion in Section \ref{sec:Introduction},
is known to cause the formation of salt sheets parallel
to the direction of the shear flow (\citep{TRAXLER2011,Radko2015,Konopliv2018,Garaud2019,Sichani2020})
in thermohaline and settling-driven convection. In the 2D simulations,
this leads to a planar interface. 

Equations \ref{eq:continuity}-\ref{eq:liquid} are specialised to
two dimensions by setting $v\equiv0$, and $\partial/\partial y\equiv0$.
We find, in general agreement with previous work on related systems, that the introduction 
of shear smooths out small-scale perturbations \citep{Konopliv2018,Garaud2019}.
We also find that increasing droplet sizes, corresponding here to
lower $Fr^{-2}$ and larger $\tau_{s,0}$, supports larger lobe sizes.
As derived in equation \ref{eq:clouds_Fr2_taus}, we have an inverse
relationship between the nondimensional parameters $Fr^{-2}$ and
$\tau_{s,0}$, which for the air-water system, and our chosen variables, appears as the dashed line in figure \ref{fig:cases}). Indeed, in general the parameters $Fr^{-2}$ and $\tau_{s,0}$ may not obey this relationship. Examples of such systems include spray drying
in the pharmaceutical industry (see e.g. \citep{Fu2012}) and heterogeneous
catalysis using microspheres \citep{Ji2010,Hu2013} that may be found
in chemical engineering industry. We therefore consider cases with
and without the inverse relationship between $Fr^{-2}$ and $\tau_{s,0}$, to mimic `cloud-like' and `non-cloud-like' flows respectively, and our simulation parameters are shown by the markers in
figure \ref{fig:cases}. We note that our derivation of the inverse relationship is restricted to small droplets. For larger droplet sizes, the settling velocity is overestimated
by the Stokes drag law, while the phase-change timescale is underestimated
if nonlinear ventilation effects are neglected \citep{Pruppacher2010}.

The simulations in 2D have an initial perturbation of the cloudy-dry
interface given in equations \ref{eq:interface_0}. Similar initial
conditions were used in RMG20, where it was found that the presence
of noise does not affect the results greatly. It was also found that
in the absence of shear, the imposed sinusoidal perturbation grows
faster than the `natural' mode that may be excited if the ratio of
the imposed wavelength to the `natural' wavelength is not too large
(see \S3 in RMG20); whereas for sufficiently large $\lambda_{x}$,
short wavelength modes are excited in addition to the growth of the
imposed wavelength. Our main results are described below.

\begin{figure}
\includegraphics[width=0.9\columnwidth]{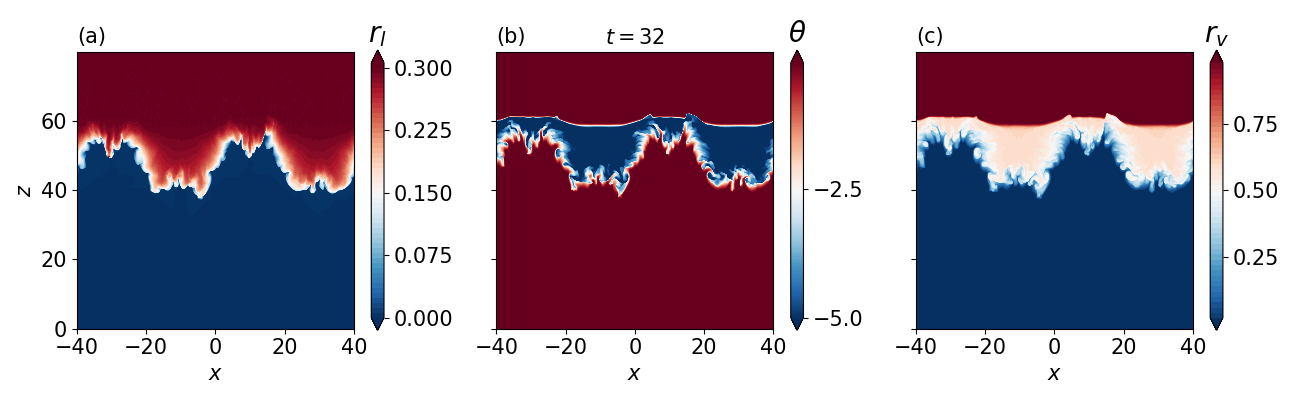}
\includegraphics[width=0.9\columnwidth]{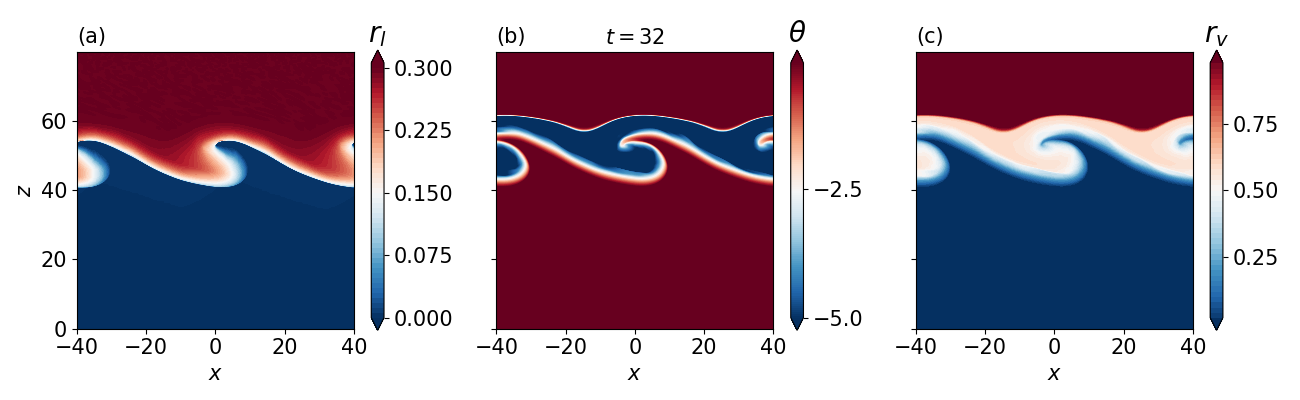}
\caption{\label{fig:S0_0.2_lambda40} Shear dampens the growth of small wavelength
instabilities. The flow parameters are $Re=1000$, $r_l^0=0.3$, $Fr^{-2}=0.1$, $\tau_{s}=20$, $z_0=60$,
with (i) $S=0$ and (ii) $S=0.2$. In each subfigure, the liquid mixing
ratio $r_{l}$ (a), the temperature (b) and the vapour mixing ratio
(c) are plotted at the time indicated. }
\end{figure}

\subsubsection{Suppression of small wavelength instabilities by shear} \label{sec:suppression}

For very low shear rates $S$, the system displays a mammatus-like
instability. For very high shear rates ($S\geq0.4$, not shown), all instabilities are suppressed
and the cloudy-dry interface becomes flat as it descends. Intermediate
shear rates, we find, suppress the short wavelength modes preferentially.
These findings are consistent with results for settling and evaporation driven
instabilities without shear, and the influence of shear on sedimentation-driven
instabilities without evaporation. First, in instabilities driven by settling and evaporation (without shear)
\citep{Ravichandran2020}, the fastest growing modes, at least in the early stages, occur 
for smaller droplet sizes and thus (in our notation here) larger $Fr^{-2}$ and smaller $\tau_s$.
We note that this is in contrast to unsheared settling-driven instabilities without evaporation, where 
the growth rates are larger for larger settling velocities \citep{Burns2012,Burns2014}.
Second, in double-diffusive and sedimentation-driven instabilities without evaporation, Konopliv {\em et al.} \citep{Konopliv2018} find, considering linearised transient optimal growth, that shear more strongly suppresses faster growing modes. In instabilities driven by settling and evaporation, these are the small-scale instabilities, and small-scale perturbations may therefore be expected to be preferentially suppressed.

Figure \ref{fig:S0_0.2_lambda40} compares the shape of the clouds with and without shear at a typical time of $t=32$. The left panels show the liquid water content, which would correspond to the visual appearance of the cloud. In all the results shown, the initial base of the cloud has been located at $z_0=60$. Water droplets have descended well below this by the time shown, and an instability has grown and evidently become nonlinear. The stabilising effect of shear on small scales is apparent. For $S=0$, small scale instabilities appear in addition to the imposed
wavelength $\lambda_x=40$. For identical initial conditions but with $S=0.2$, the small scale instabilities are suppressed and only the $\lambda_x=40$ mode is seen. Moreover, as is to be expected, the structures are now tilted by the shear. The middle panels show the temperature profiles. Evaporative cooling is evident (blue regions in the figure) in the portion of the cloud where droplets have descended from the original cloud base. In this region, the water vapour content (right panels) is higher than that of dry air, as expected from evaporation. There is a broad correspondence in the shape of structures displayed by the three physical quantities, although they are different in detail. We will see that in three dimensions, there are greater differences between the structures displayed by the three scalars.

\begin{figure}
    \centering
    \includegraphics[width=0.5\columnwidth]{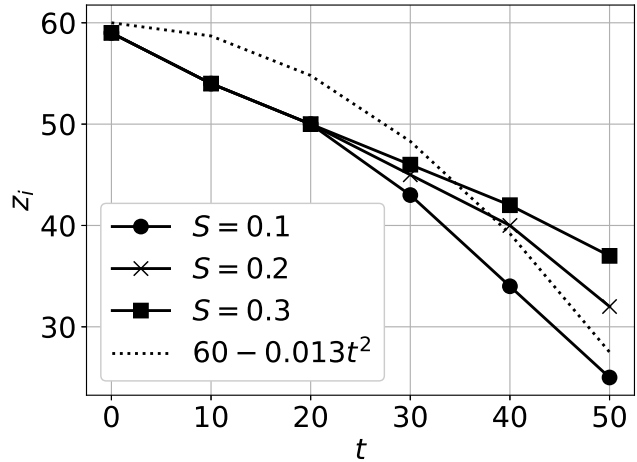}
    \caption{The location of the cloudy-dry interface, $z_i$, as a function of time as the shear rate $S$ is varied, with $Fr^{-2}=0.1$, $\tau_{s,0}=20$, $\lambda_x=20$. Shear suppresses the settling and evaporation-driven instability, leading to smaller mixing rates.}
    \label{fig:zi_vs_S}
\end{figure}
The rate at which the cloudy-dry interface moves in the $-z$ direction is a measure of the rate of mixing by the flow. Since the region $z<z_0$ is initially dry ($r_v=0$) we locate this interface at $z=z_i$ at which the horizontally averaged vapour mixing ratio $\bar{r}_v(z)$ takes on a threshold value of $0.1$. In figure \ref{fig:zi_vs_S}, we plot the location of this interface for three values of shear. At early times, the instability develops similarly in the three cases, and mixing levels are similar.  But at later times, larger shear suppresses the development of the instability structures and retards the mixing.
\begin{figure}
\includegraphics[width=0.9\columnwidth]{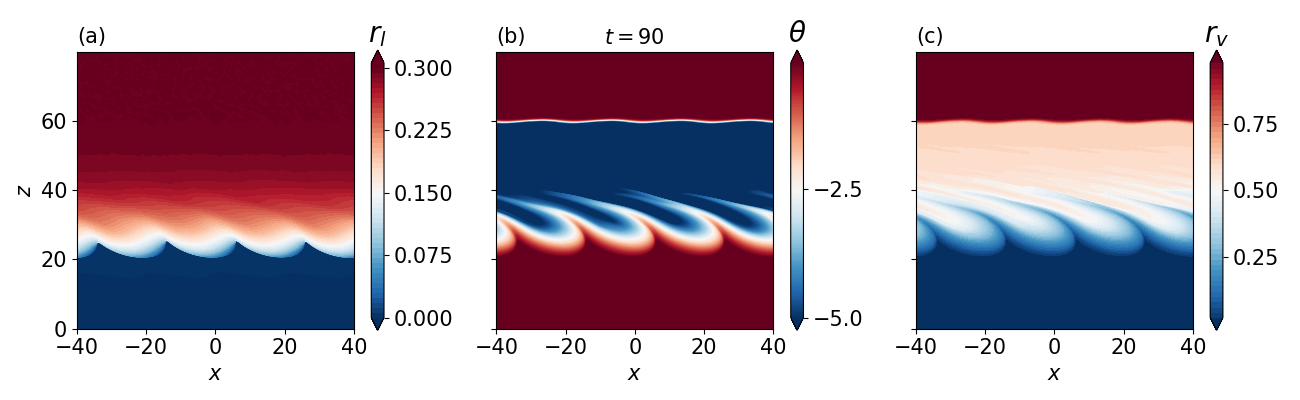}
\includegraphics[width=0.9\columnwidth]{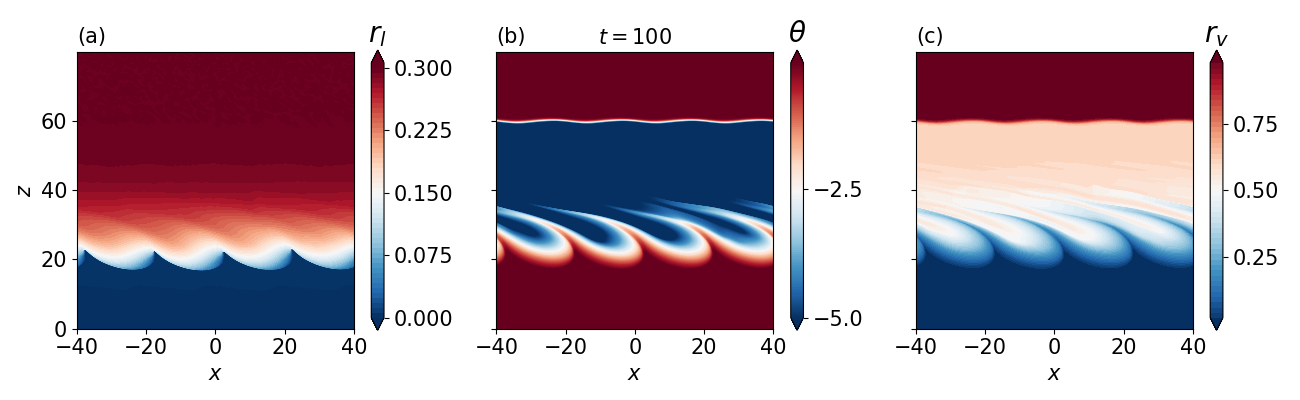}
\caption{\label{fig:resonance} For sufficiently large shear, the tendency of the instability
to grow and the tendency of the shear to suppress growth balance each other, leading to lobes that
maintain their size and approximately their shape as they descend. The parameters are
$Fr^{-2}=0.04$, $\tau_s=50$, $S=0.1$, $z_0=60$, $\lambda_x=20$, and the figures are plotted at the times
indicated. }
\end{figure}

\begin{figure}
    \centering
    \includegraphics[width=0.8\columnwidth]{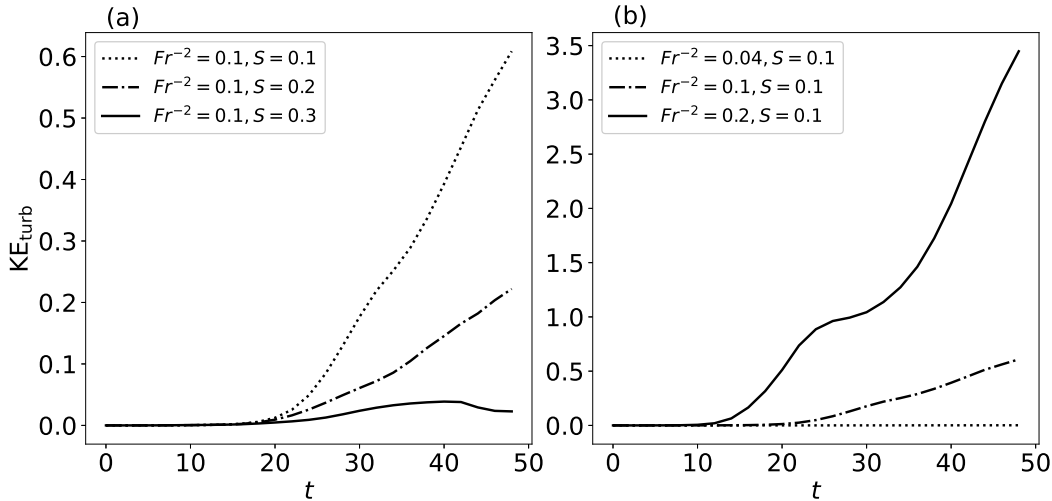}
    \caption{\label{fig:KE_turb} \tb{The turbulent kinetic energy (see text for definition) in the `cloud-like' cases as a function of time for (a) different $S$ with $Fr^{-2}=0.1$ and (b) different $Fr^{-2}$ with $S=0.1$, with $\lambda_x=20$. Note that the curve for $Fr^{-2}=0.1, S=0.1$ appears in both subfigures. }}
\end{figure}

\begin{figure}
    \includegraphics[width=0.4\columnwidth]{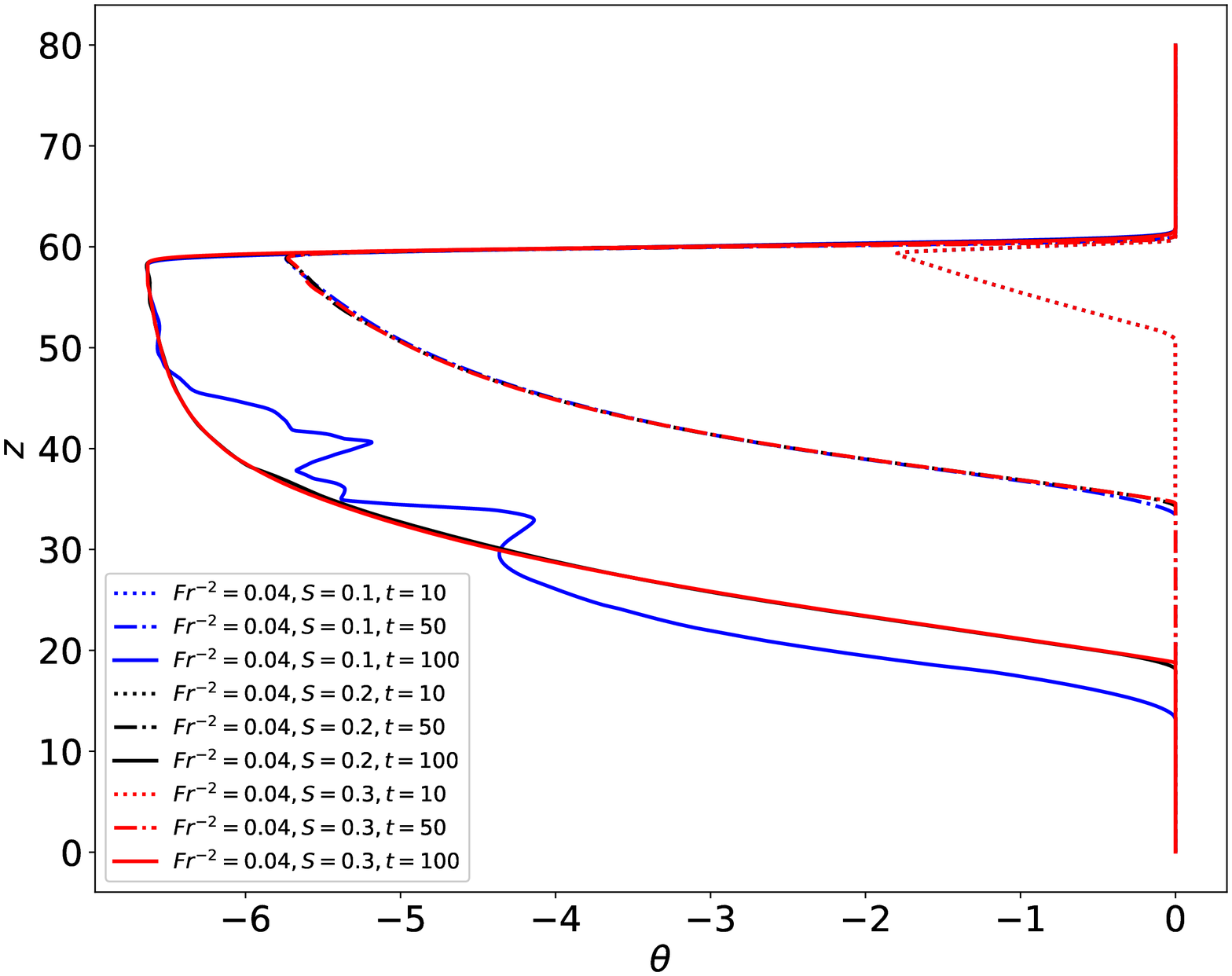}    \includegraphics[width=0.4\columnwidth]{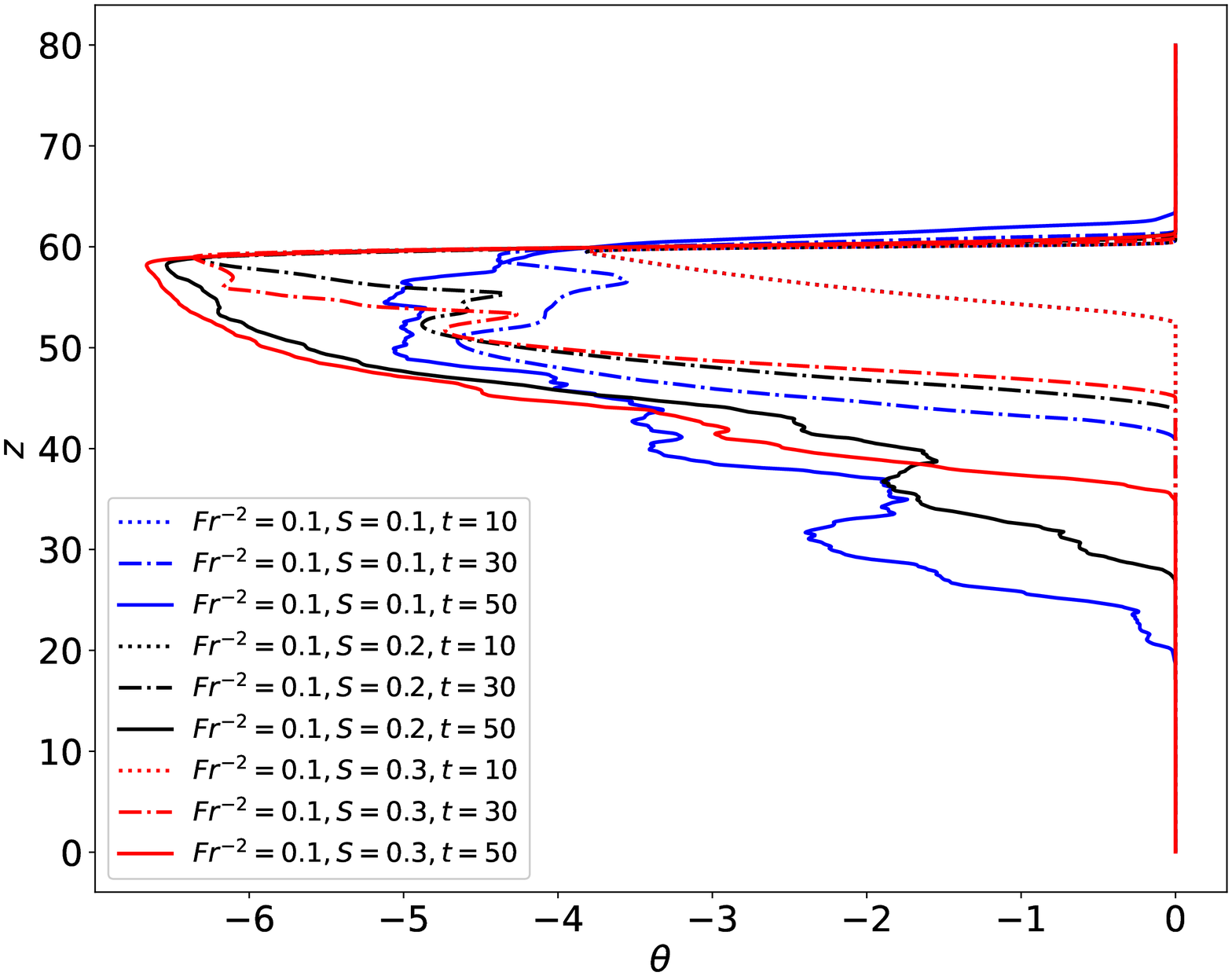}    \includegraphics[width=0.4\columnwidth]{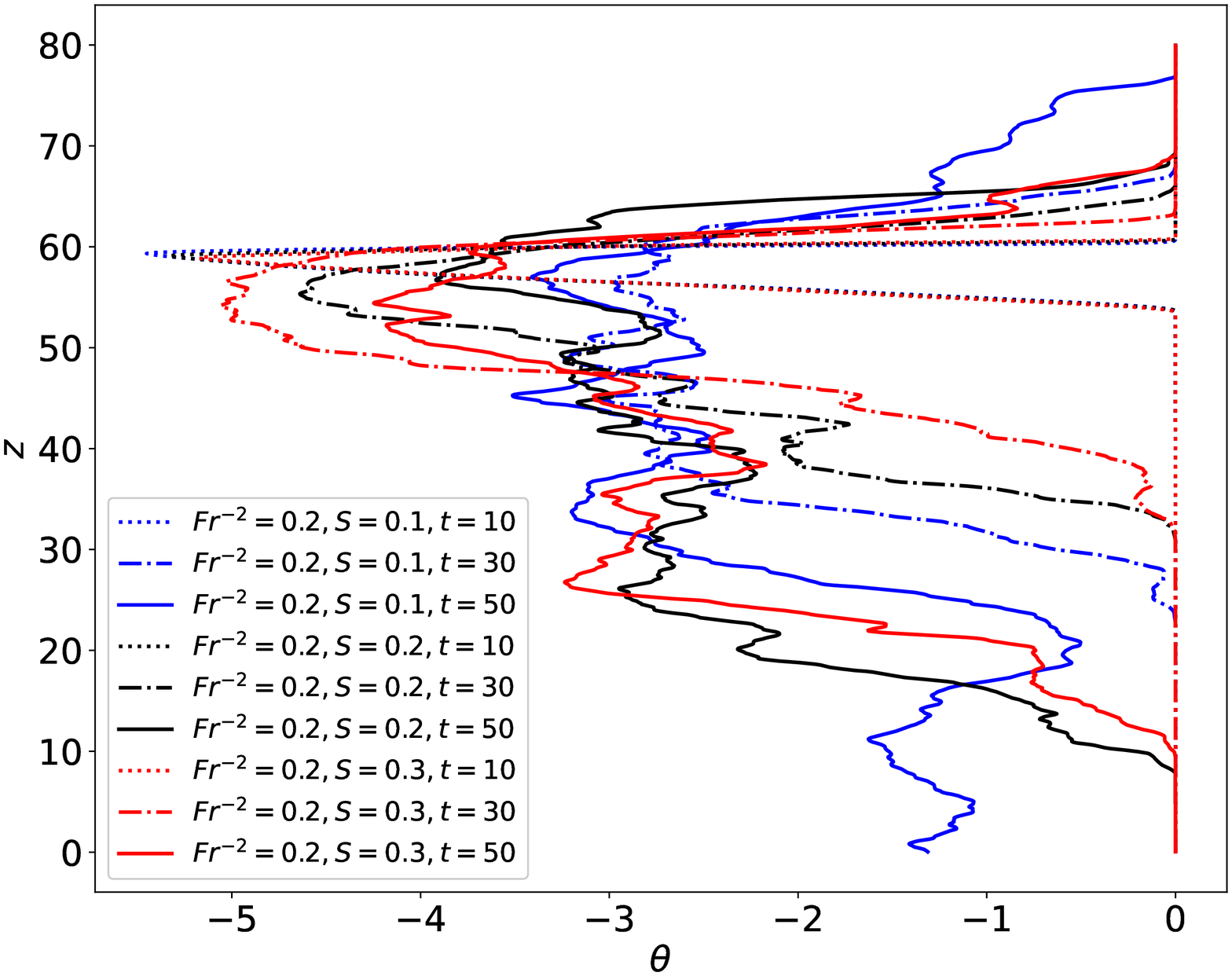}
    \caption{\label{fig:thbar_vs_z} \tb{Horizontally averaged profiles of the temperature $\theta$, plotted at three times and three values of the shear rate $S$ (a) $Fr^{-2}=0.04$, (b) $Fr^{-2}=0.1$, (c) $Fr^{-2}=0.2$ with $\lambda_x=20$. In each case, shear suppresses the instability, leading to smaller mixed layer depths. }}
\end{figure}

\tb{The suppression of flow instabilities by shear is also evidenced in figure \ref{fig:KE_turb}, where we plot the domain-averaged turbulent kinetic energy\\
\[
\mathrm{KE}_\mathrm{turb} = \mathrm{KE} - \mathrm{KE}(t=0),
\]
\tb{where the base value due to the background shear (the kinetic energy at $t=0$) is subtracted.} For each combination of `cloud-like' $Fr^{-2}$ and $\tau_s$, we see that larger shear leads to smaller kinetic energy.}
\tb{We examine the evolution of the mixed layers by plotting the horizontally averaged temperature excess in figure \ref{fig:thbar_vs_z}, showing that the rate at which the mixed layers grow is inversely related to the shear rate $S$. We also note that the mixed layers grow only for $z<z_0$ for large droplet sizes (small $Fr^{-2}=0.04,0.1$) while for $Fr^{-2}=0.2$, the turbulent flow penetrates the stably stratification at $z=z_0$ and thus the mixed layer grows in both directions from $z=z_0$. Profiles of other variables show similar behaviour. }

At significant shear rate and for large droplet sizes (i.e., smaller $Fr^{-2}$ and larger $\tau_s$ than seen in figure \ref{fig:S0_0.2_lambda40}), the instabilities grow rather slowly, so the sheared lobes maintain their appearance as they descend for tens of flow units, as seen in figure \ref{fig:resonance}. The lobe sizes that can be maintained in this way are larger for higher shear rates. We remark that the structures seen are not Lagrangian objects that are descending. They are constantly being renewed by evaporation, and regeneration of the instability while the shape is being maintained. For small droplet sizes on the other hand, the flow becomes turbulent and structure shapes are not maintained over significant durations. This may be surmised from figures \ref{fig:zi_vs_S} and from \ref{fig:zi_vs_tau_Fr} discussed below, both of which show the influence of different parameters on the mixing rate.

\subsubsection{Effect of deviation from `cloud-like' parameters}\label{sec:cloud-like}

\begin{figure}
\includegraphics[width=0.9\columnwidth]{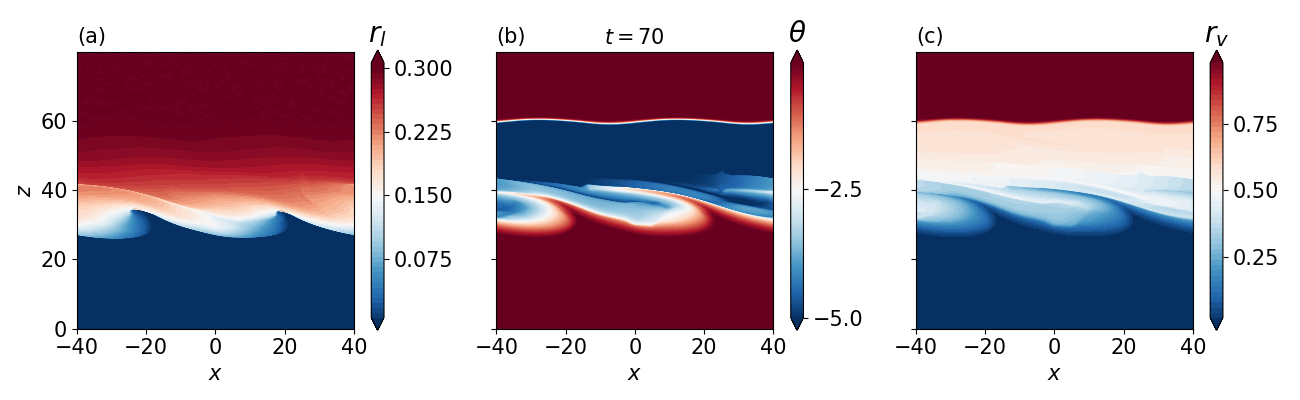}
\includegraphics[width=0.9\columnwidth]{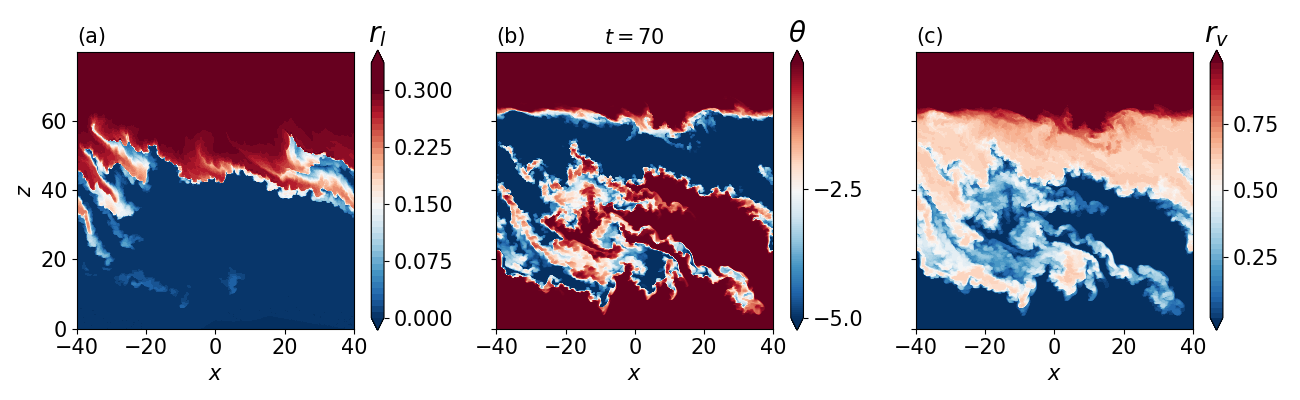}
\caption{\label{fig:non_cloud_like} With all other parameters held fixed, slower evaporation than `cloud-like', i.e.,  $\tau_s=50$ (a) leads
to a lobe-like instability, while faster evaporation than `cloud-like', i.e.,  $\tau_s=10$ (b) leads to small scale instabilities
which lead to much greater mixing and larger fluxes of vapour and energy (although not liquid; see text).
The parameters are $Fr^{-2}=0.1$,  $S=0.25$, $z_0=60$, $\lambda_x=40$, and the figures are plotted at the times
indicated.}
\end{figure}
\begin{figure}
    \centering
    \includegraphics[width=0.9\columnwidth]{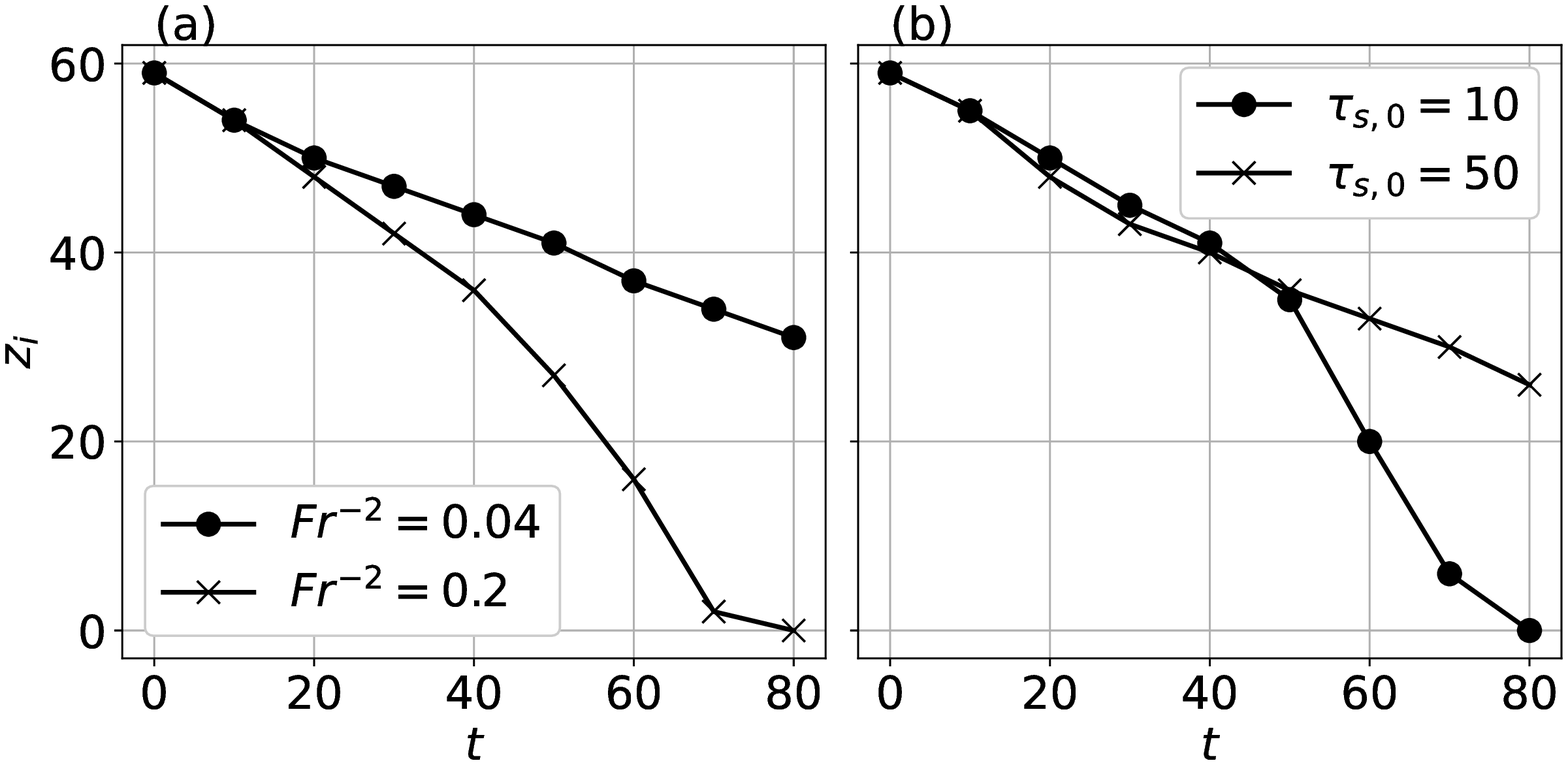}
    \caption{The location of the cloudy-dry interface $z_i$, defined as in figure \ref{fig:zi_vs_S}, as a function of time for (a) $\tau_{s,0}=20$ and two values of $Fr^{-2}$, and (b) $Fr^{-2}=0.1$, and two values of $\tau_{s,0}$, showing that the mixing is faster for smaller $\tau_s$ and larger $Fr^{-2}$. The other parameters are $S=0.25$ and $\lambda_x=20$.
    \label{fig:zi_vs_tau_Fr}}
\end{figure}
As discussed in section \ref{sec:Setup}, `cloud-like' cases have
$Fr^{-2}\tau_{s,0}\approx2$ (see figure \ref{fig:cases}), while this relation may not hold in other systems of interest with a settling and reacting scalar component. To ask how a departure from cloud-like physical parameters will affect the instability, we explore in figure \ref{fig:non_cloud_like}, the consequence of keeping $Fr^{-2}=0.1$ while changing $\tau_{s,0}$, i.e., moving along the horizontal in figure \ref{fig:cases}. The `cloud-like' case ($Fr^{-2}=0.1$, $\tau_{s,0}=20$) is expected to be similar to the lower panel of figure \ref{fig:S0_0.2_lambda40}.  Smaller droplets (lower $\tau_{s,0}$) evaporate faster, leading to larger scalar fluxes of vapour and energy, and lower liquid water content, given that it is the evaporation of the liquid that is the source of kinetic energy in the flow. Interestingly therefore, the liquid in this case occupies a smaller region below the `cloud' as compared to the temperature and vapour fields, which means that the visual appearance of such a flow would not be indicative of the other scalars. The structures in temperature and vapour are less well-formed, and more chaotic in appearance than in a cloud-like case. In general, for the same $Fr^{-2}$,
larger values of $\tau_{s,0}$ lead to (lobe-like) instabilities of larger wavelengths,
while smaller values of $\tau_s$ lead to smaller-scale instabilities;
conversely, for the same $\tau_{s,0}$, larger values
of $Fr^{-2}$ (i.e. stronger buoyancy) result in instabilities of
smaller wavelengths, while weak buoyancy (smaller $Fr^{-2}$) leads to larger length-scale instabilities. We see thus that flows made up of larger droplets are more likely to take on an asperitas-like appearance, and also be longer-lived. Further evidence for this is given in figure \ref{fig:zi_vs_tau_Fr}. Here we plot the horizontally-averaged vertical location $z_i$ of the `cloudy'-dry interface as $Fr^{-2}$ and $\tau_{s,0}$ are varied. A slope of $-1$ on these graphs would indicate a mixing layer growing at the same rate as the initial particle settling velocity. The behaviour we see may be explained by noting that both smaller $\tau_s$ and larger $Fr^{-2}$ are representative of smaller droplets, which evaporate faster for a given $r_l$, leading to higher sub-cloud cooling, and therefore a greater role for buoyancy. At higher levels of buoyancy, observations in settling-driven two-component systems without phase change show \citep[e.g.][]{Burns2014} that the velocity of the interface is higher, and this is consistent with what we see. The additional physics in our case is that we do not impose the buoyancy forcing externally, but it gets generated as a result of the thermodynamics of phase change.

\subsection{3D simulations} \label{sec:3D}

\begin{figure}
\includegraphics[width=0.5\columnwidth]{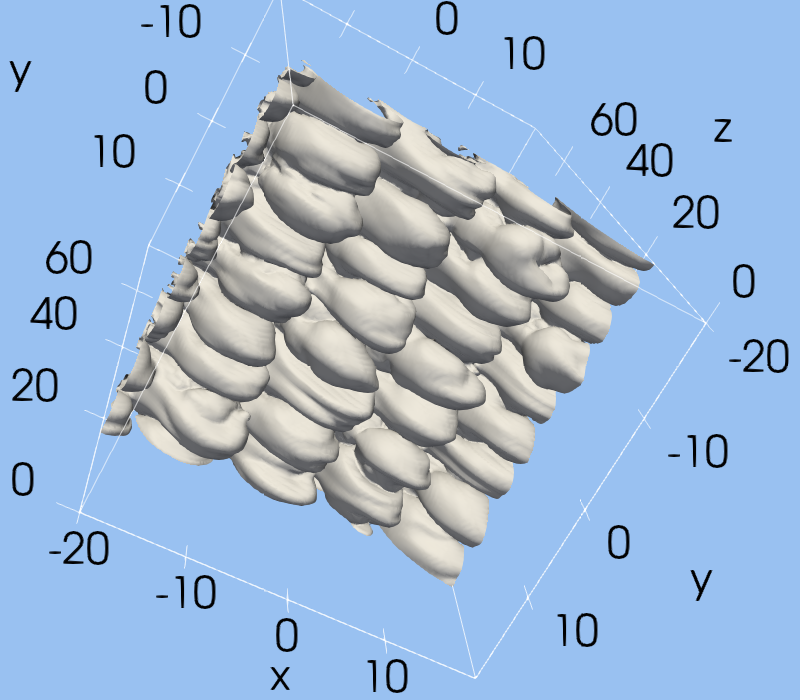}\includegraphics[width=0.5\columnwidth]{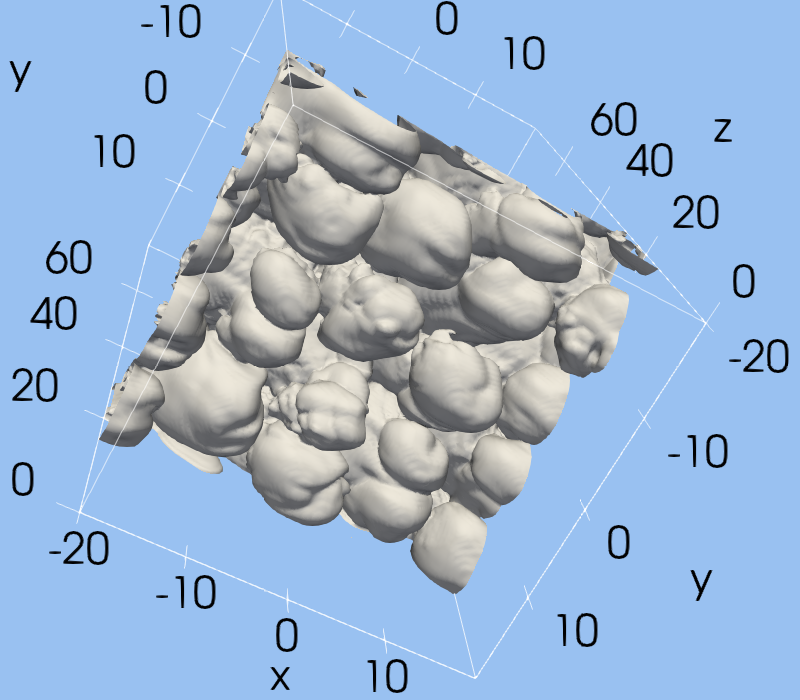}
\includegraphics[width=0.5\columnwidth]{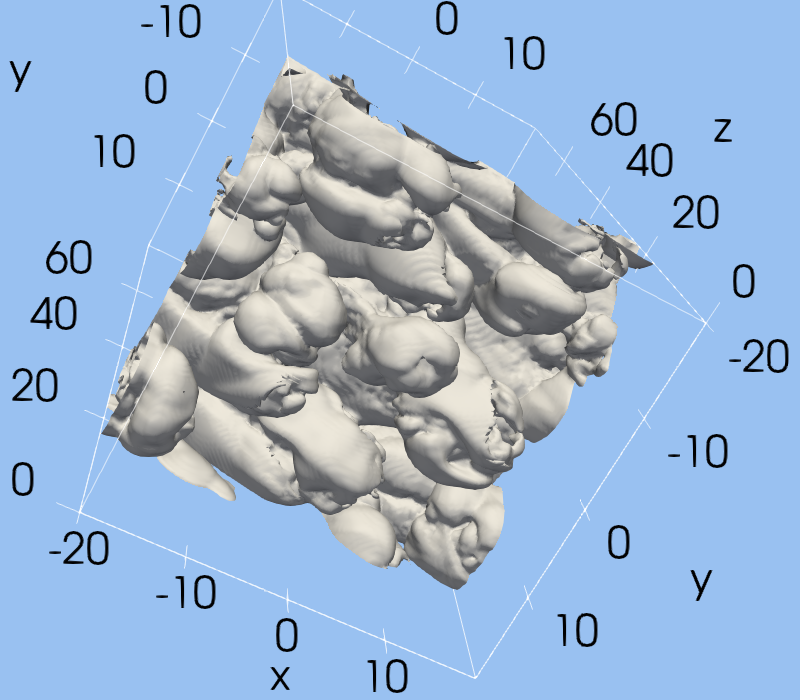}\includegraphics[width=0.5\columnwidth]{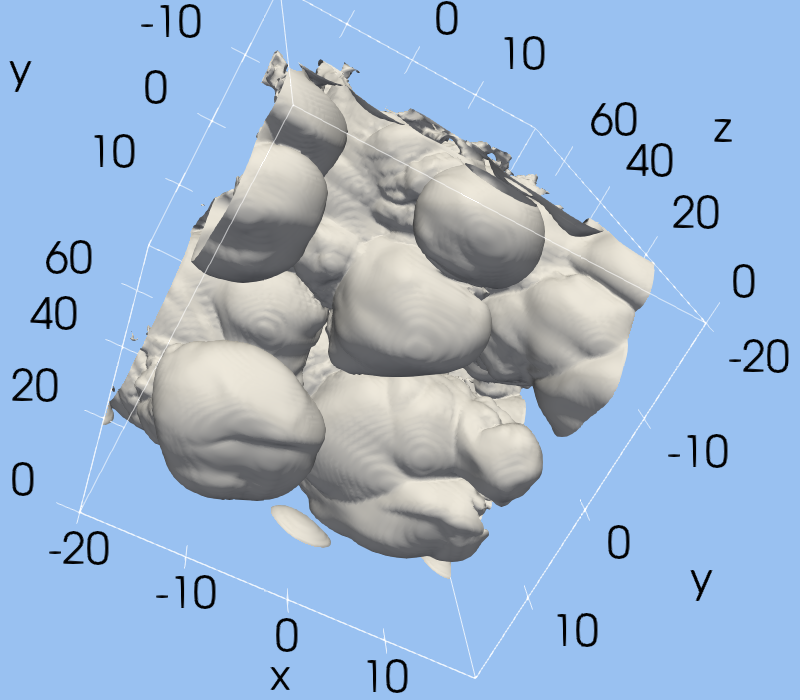}
\caption{\label{fig:3D_S0.05} The growth of the sheared lobes with $S=0.05$,
$\lambda_{x}=20$ and (a) $\lambda_{y}=5$, (b) $\lambda_{y}=10$,
(c) $\lambda_{y}=20$, (d) $\lambda_{y}=40$. The other parameters
are $Fr^{-2}=0.04$ and $\tau_{s}=50$. The iso-surfaces are plotted
at $r_{l}=0.01$ and time $t=80$. The lobes in (d) have reached $z\approx10$ and
developed while those in (a) have only reached $z=20$. C.f. figure
\ref{fig:3D_S0.1} with all the same parameters except a larger shear
rate. }
\end{figure}
A systematic analysis of the growth rate of the instability in 3D as the
horizontal wavelengths are varied will be presented elsewhere, since the main aim at present is to bring the physics to light. For this purpose it is sufficient to present results from cases with $Fr^{-2}=0.04$, $\tau_{s}=50$
(`cloud-like'), while varying the shear rate $S$ and the initial
perturbation wavelengths $\lambda_{x}$ and $\lambda_{y}$. Due to
computational constraints, the simulations were performed with half
the grid resolution (i.e. twice the grid spacing) as the 2D simulations,
and with horizontal domains of $40\times40$.

As discussed in section \ref{sec:Introduction}, shear leads to the
formation of `salt sheets' in double-diffusive or sediment-driven
convection. For small shear rates, the mammatus-lobe instability results.
For large shear rates (see equation \ref{eq:shear_0}), the interface
evolves to become homogeneous in the $x-$direction, leading to `liquid
sheets'. For intermediate shear, as in 2D, the nature of the lobes
that form is determined by the competition between the growth rate
and the homogenising effect of shear. 

The central effect of shear is as follows.
RMG20 found that the growth rates of linear instabilities driven by settling
and evaporation is larger for smaller droplet sizes (of smaller settling velocities), which lead to
 smaller instability wavelengths. However,
as we have seen, the homogenising effects of shear suppress instabilities
of the small wavelengths. As a result, when shear is non-zero, larger
wavelengths have larger growth rates. This is seen in figure \ref{fig:3D_S0.05} which shows isosurfaces at $r_l=0.01$ at a typical time. A small value of liquid water content is chosen to plot the isosurfaces, so that the images may correspond closely to the visual appearance of the cloud. It is seen that (a) larger shear rates lead to smaller growth rates, other things being held
constant; and (b) larger wavelengths lead to more developed lobes
at a given time for the same shear. We also note that the effect of
the shear is most obviously seen in the wavenumber of the instability
in the $y-$direction, which is twice as large as the imposed $y-$wavenumber.

\begin{figure}
\includegraphics[width=0.5\columnwidth]{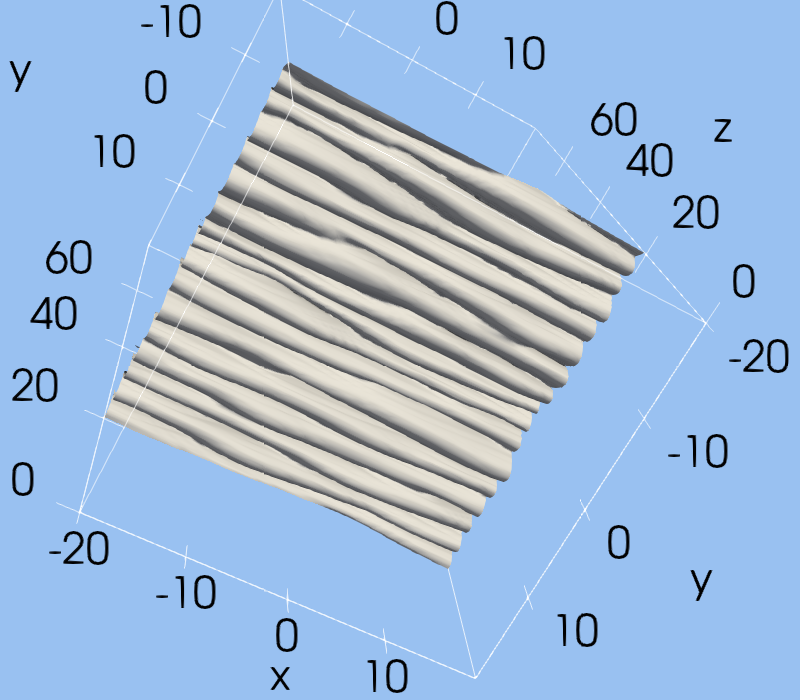}\includegraphics[width=0.5\columnwidth]{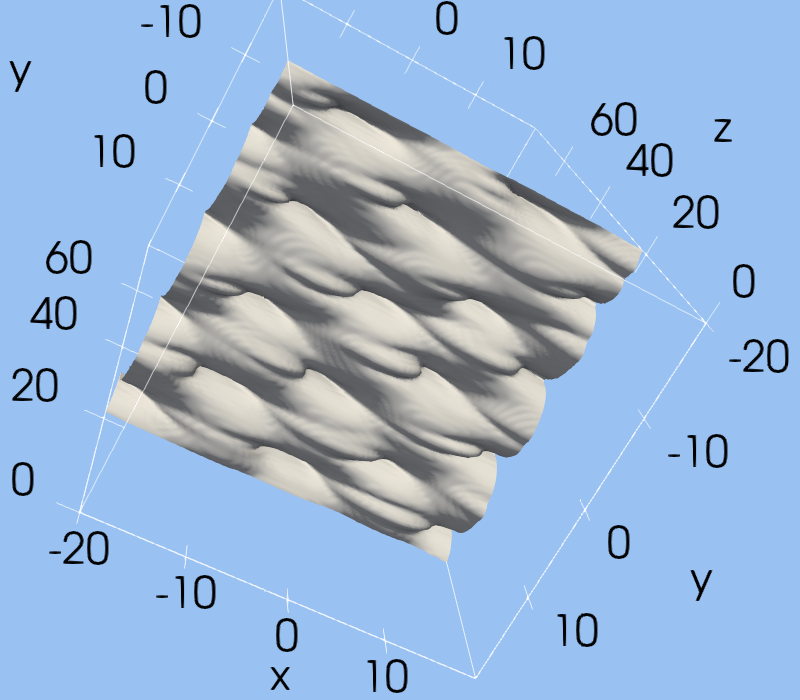}
\includegraphics[width=0.5\columnwidth]{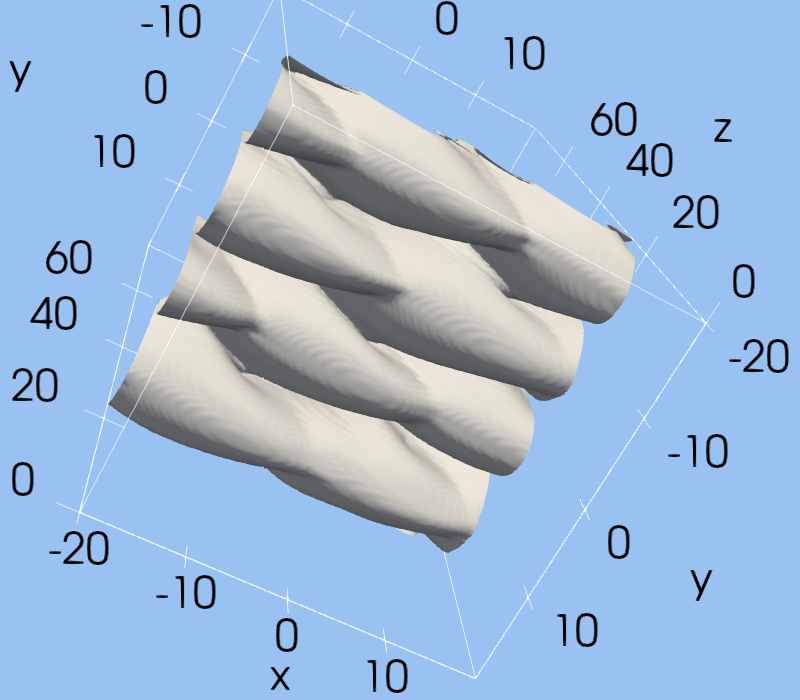}\includegraphics[width=0.5\columnwidth]{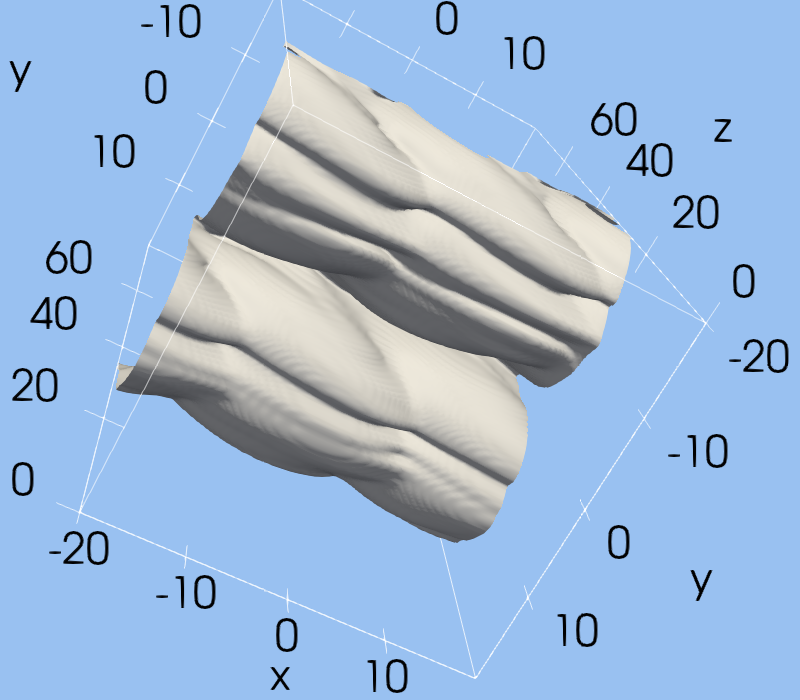}
\caption{\label{fig:3D_S0.1}The growth of the sheared lobes at the same time $t=80$, and with the same
parameters as figure \ref{fig:3D_S0.05} except with a larger shear
rate $S=0.1$. The iso-surfaces are plotted at $r_{l}=0.01$ as in
figure \ref{fig:3D_S0.05}, and show that the instability is more pronounced
for when the $x$ and $y$ wavelengths of the perturbations are large and comparable to each other.}
\end{figure}

\begin{figure}
\includegraphics[width=1\columnwidth]{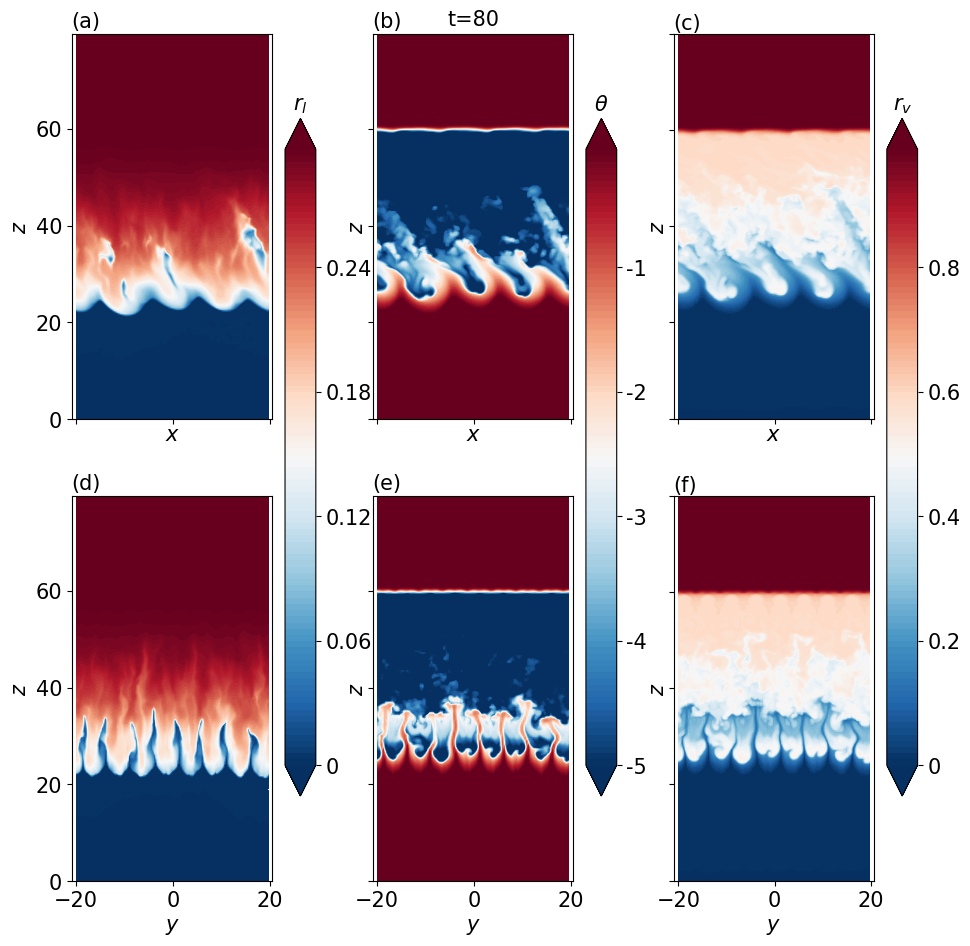}

\caption{\label{fig:3D_S0.05_lamz5xyxz} Two views of the liquid mixing
ratio $r_{l}$ (a,c), the temperature $\theta$ (b,d) and the vapour
mixing ratio $r_{v}$ (c,f) on (a-c) $x-z$ planes at $y=0$ and (d-f)
$y-z$ planes at $x=0$. The parameters are $S=0.05$, $Fr^{-2}=0.04$,
$\tau_{s}=50$, $\lambda_{x}=20$, $\lambda_{y}=5$, as in figure \ref{fig:3D_S0.05}(a). C.f. figure \ref{fig:3D_S0.1_lamz5_xyxz}
with all the same parameters except a larger shear rate.  }
\end{figure}

\begin{figure}
\includegraphics[width=1\columnwidth]{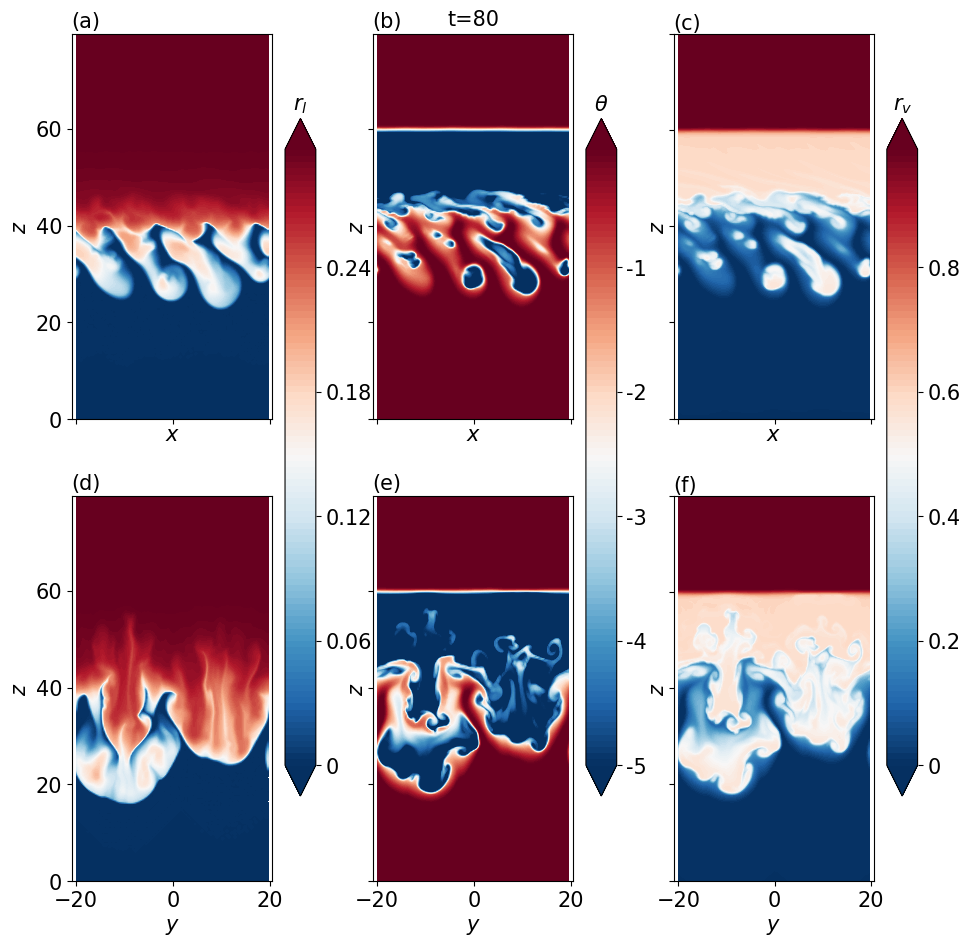}

\caption{\label{fig:3D_S0.05_lamz40_xyxz} Two cross-sections of the simulations shown in figure \ref{fig:3D_S0.05}(d) of the liquid mixing
ratio $r_{l}$ (a,c), the temperature $\theta$ (b,d) and the vapour
mixing ratio $r_{v}$ (c,f) on (a-c) $x-z$ planes at $y=0$ and (d-f)
$y-z$ planes at $x=0$. The parameters are $S=0.05$, $Fr^{-2}=0.04$,
$\tau_{s}=50$, $\lambda_{x}=20$, $\lambda_{y}=40$. C.f. figure
\ref{fig:3D_S0.1_lamz40_xyxz} with all the same parameters except
a larger shear rate.}
\end{figure}

\begin{figure}
\includegraphics[width=1\columnwidth]{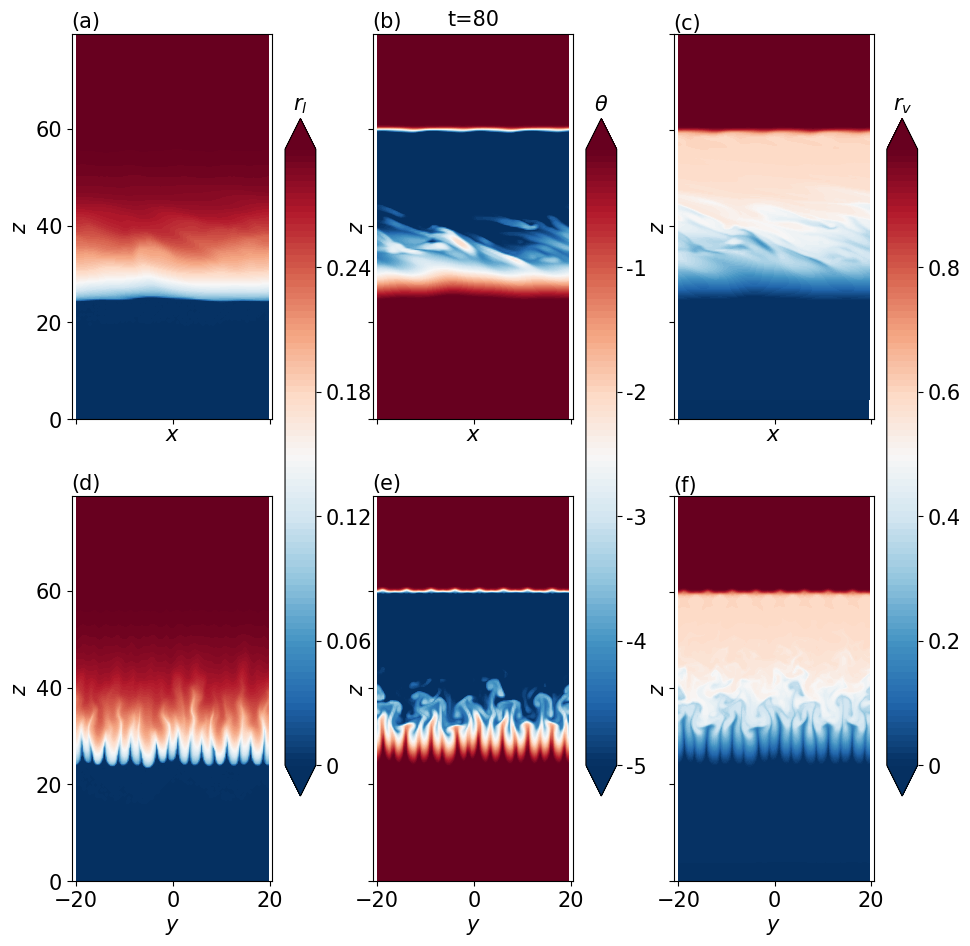}

\caption{\label{fig:3D_S0.1_lamz5_xyxz} Cross-sections of the liquid mixing
ratio $r_{l}$ (a,c), the temperature $\theta$ (b,d) and the vapour
mixing ratio $r_{v}$ (c,f) on (a-c) $x-z$ planes at $y=0$ and (d-f)
$y-z$ planes at $x=0$, for the same parameters as figure \ref{fig:3D_S0.05_lamz5xyxz}
except with a larger shear rate $S=0.1$. At late times, the flow
is homogeneous in the $x-$direction.}
\end{figure}

\begin{figure}
\includegraphics[width=1\columnwidth]{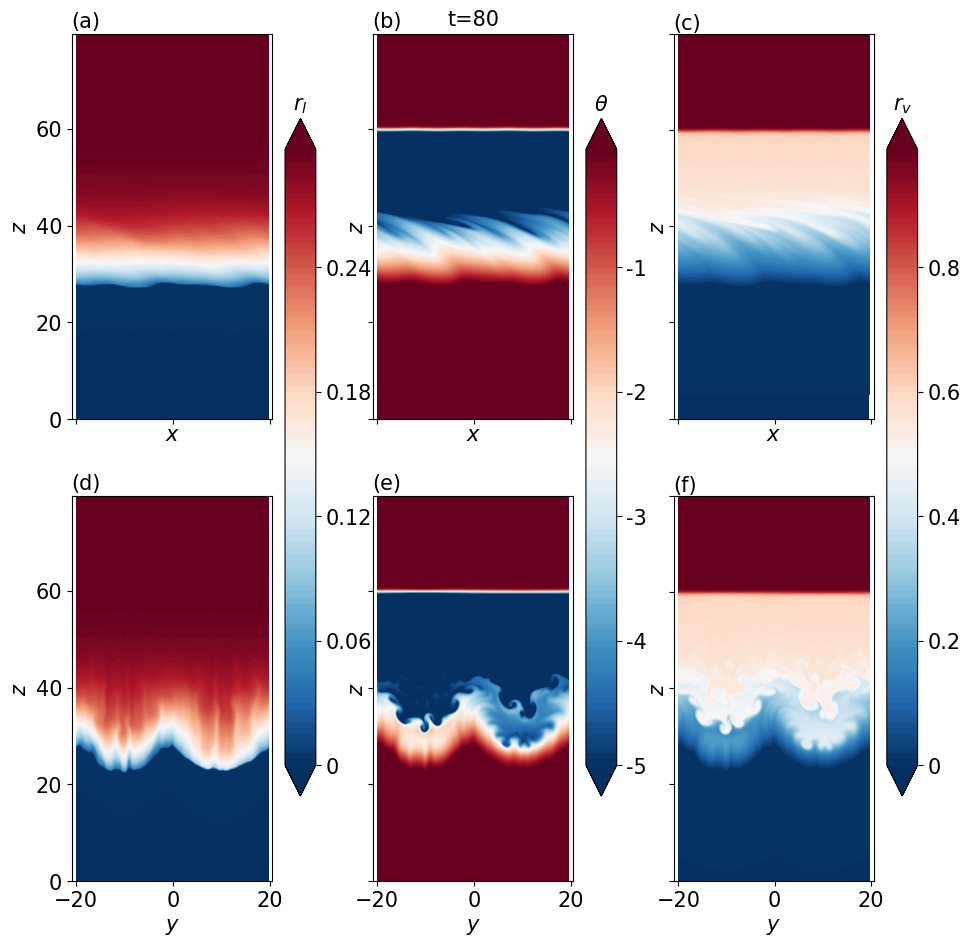}

\caption{\label{fig:3D_S0.1_lamz40_xyxz}Cross-sections of the liquid mixing
ratio $r_{l}$ (a,c), the temperature $\theta$ (b,d) and the vapour
mixing ratio $r_{v}$ (c,f) on (a-c) $x-z$ planes at $y=0$ and (d-f)
$y-z$ planes at $x=0$, for the same parameters as figure \ref{fig:3D_S0.05_lamz40_xyxz}
except with a larger shear rate $S=0.1$. At late times, the flow
is homogeneous in the $x-$direction.}
\end{figure}

Figures \ref{fig:3D_S0.05} and \ref{fig:3D_S0.1} and the corresponding
cross-sections of the flow variables, plotted in figures \ref{fig:3D_S0.05_lamz5xyxz}-\ref{fig:3D_S0.1_lamz40_xyxz},
show that the equivalent of `salt-sheets' are formed for the larger
shear rate $S=0.1$. Thus, the shear rate required to homogenise the
flow along the direction of the shear flow is smaller in 3D than in
2D. A revealing aspect of the 3D simulations is that high shear organises the structures into elongated formations rather than the blob-like appearance which was seen in mammatus clouds. Moreover, when the perturbation wavelengths in the two horizontal directions are very different (figures \ref{fig:3D_S0.05} (a) and \ref{fig:3D_S0.1} (a)), higher shear leads to thinner and feebler structures. In the light of figures \ref{fig:3D_S0.05} and \ref{fig:3D_S0.1}, we may return to figure \ref{fig:asperitas_web_pics}. It is apparent that suitably designed initial conditions could lead to numerical formations of the kind seen in real asperitas clouds. The effect of shear, and of the wavelength of the dominant perturbation,  on the 3D instability structures is made more evident in the cross-sectional views shown in figures \ref{fig:3D_S0.05_lamz5xyxz} - \ref{fig:3D_S0.1_lamz40_xyxz}. First we notice that there is far less structure in the plane of the shear (the $x-z$ plane) than in the $y-z$ plane. Structures are broadly aligned in the streamwise direction, and this is reminiscent of streamwise streaks in standard Couette flow which are generated by the non-modal growth of perturbations. Second, the structures look remarkably periodic despite the flow being quite turbulent, which is evident from the plots of the vapour field $r_v$. Since descent dominates the liquid field, we do not see much turbulence here.

\section{Conclusions and Outlook \label{sec:Conclusion}}
We have studied the effects of shear on the instability driven by the evaporation of a scalar component that settles under gravity. In both two and three dimensions, shear suppresses the
development of the settling and evaporation driven instability, in analogy with similar findings for 
fingering and sedimentation driven instabilities. We showed, additionally, that small-scale instabilities are
suppressed preferentially by the shear. We argued that this is because shear 
more strongly suppress the fastest growing sedimentation-driven modes which, when evaporation is also present,
are of small-scale. We then showed the existence in 2D of a balance between shear and the instability which
preserves the size and shape of descending lobes. Lastly, we showed that when $Fr^{-2}$ and $\tau_{s,0}$ are
varied independently, smaller $Fr^{-2}$ and larger $\tau_{s,0}$ lead to lobe-like instabilities and smaller scalar fluxes.

In 3D, we found that the instabilities, at least when the initial perturbation wavelength is externally imposed, rapidly
lead to the sheet-like structures previously observed in the sedimentation-driven and double-diffusive fingering regimes.
For small shear rates, we showed that distorted lobe-like structures resembling asperitas clouds can be generated. For large shear rates, the instabilities become homogeneous in the direction of the shear flow, producing a closer resemblance to clouds in figure \ref{fig:asperitas_web_pics}(b).
Since shear almost always exists in the troposphere (as also evidenced in the few known measurements associated with
asperitas observations), this suggests shear-influenced settling and evaporation as a possible mechanism for the formation
of asperitas clouds. 

The mechanism laid down here makes specific predictions about why asperitas clouds are rarer \cite{GilesHarrison2017} 
than mammatus clouds. The formation of mammatus clouds by settling and evaporation requires sufficiently large droplet sizes as well as
sufficiently large liquid water content. These criteria also apply to the formation of asperitas clouds. We have shown that shear is an important additional ingredient in the formation of asperitas-like clouds. And that when the shear rates are either too small (and the clouds resemble mammatus clouds) or too large (when the different lobes merge in the
direction of shear and the structures are unable to grow to significant depth), asperitas clouds may not form. In the case of mammatus clouds, RMG20 found that the results were qualitatively the same in 2D and 3D, and blob-like protruberances are seen in both. In asperitas clouds, however, 3D simulations are necessary to distinguish between blob-like and undulating formations. It is shown that the latter are more likely at higher shear.

We note that our simulations are highly idealised and the caveats laid down in RMG20 apply here. In particular, our
Reynolds number of $Re=1000$ is  significantly smaller than realistic values.  We have applied a given wavenumber in each direction in our simulations whereas a real asperitas cloud would obviously display a combination of these. In fact it would be an interesting exercise to emulate asperitas shapes seen in the atmosphere by prescribing initial perturbations which contain a spectrum of wavelengths. Furthermore, as with mammatus clouds, the settling and evaporation-driven mechanism may not explain all occurrences of asperitas
clouds. 

Some avenues for future work are evident. We mention two here.\\
1) The optimal combination of $Fr^{-2}$ and $\tau_s$, if they can be varied independently, that maximises the flux of a given scalar quantity (with and without externally imposed shear, and in 2D and 3D) would be relevant for optimising certain processes in the chemical engineering industry. Such results for settling-driven and double-diffusive convection have been obtained, e.g. in \cite{Burns2014}. Given our focus on the elucidating the mechanism and on the visual patterns formed by the instability, a systematic parametric study is beyond the scope of this work. However, the results in figures \ref{fig:zi_vs_S} and \ref{fig:zi_vs_tau_Fr} may be considered initial findings in that direction. We believe our formulation of settling and evaporation-driven, shear modulated, convection in terms of the three nondimensional parameters $Fr^{-2}$, $\tau_s$ and $S$ has laid the foundation for this effort.
\\
2) Double-diffusive and sedimentation driven instabilities are known to lead to `staircases', i.e., to alternating horizontal layers containing, and devoid
of, the scalar component(s) \citep[see, e.g.][]{TRAXLER2011,Radko2012,Ouillon2020}. It would be of great interest to find whether such staircases
are possible in a system where one scalar component can evaporate.


\begin{thebibliography}{34}%
\makeatletter
\providecommand \@ifxundefined [1]{%
 \@ifx{#1\undefined}
}%
\providecommand \@ifnum [1]{%
 \ifnum #1\expandafter \@firstoftwo
 \else \expandafter \@secondoftwo
 \fi
}%
\providecommand \@ifx [1]{%
 \ifx #1\expandafter \@firstoftwo
 \else \expandafter \@secondoftwo
 \fi
}%
\providecommand \natexlab [1]{#1}%
\providecommand \enquote  [1]{``#1''}%
\providecommand \bibnamefont  [1]{#1}%
\providecommand \bibfnamefont [1]{#1}%
\providecommand \citenamefont [1]{#1}%
\providecommand \href@noop [0]{\@secondoftwo}%
\providecommand \href [0]{\begingroup \@sanitize@url \@href}%
\providecommand \@href[1]{\@@startlink{#1}\@@href}%
\providecommand \@@href[1]{\endgroup#1\@@endlink}%
\providecommand \@sanitize@url [0]{\catcode `\\12\catcode `\$12\catcode
  `\&12\catcode `\#12\catcode `\^12\catcode `\_12\catcode `\%12\relax}%
\providecommand \@@startlink[1]{}%
\providecommand \@@endlink[0]{}%
\providecommand \url  [0]{\begingroup\@sanitize@url \@url }%
\providecommand \@url [1]{\endgroup\@href {#1}{\urlprefix }}%
\providecommand \urlprefix  [0]{URL }%
\providecommand \Eprint [0]{\href }%
\providecommand \doibase [0]{http://dx.doi.org/}%
\providecommand \selectlanguage [0]{\@gobble}%
\providecommand \bibinfo  [0]{\@secondoftwo}%
\providecommand \bibfield  [0]{\@secondoftwo}%
\providecommand \translation [1]{[#1]}%
\providecommand \BibitemOpen [0]{}%
\providecommand \bibitemStop [0]{}%
\providecommand \bibitemNoStop [0]{.\EOS\space}%
\providecommand \EOS [0]{\spacefactor3000\relax}%
\providecommand \BibitemShut  [1]{\csname bibitem#1\endcsname}%
\let\auto@bib@innerbib\@empty
\bibitem [{\citenamefont {Ravichandran}\ \emph {et~al.}(2020)\citenamefont
  {Ravichandran}, \citenamefont {Meiburg},\ and\ \citenamefont
  {Govindarajan}}]{Ravichandran2020}%
  \BibitemOpen
  \bibfield  {author} {\bibinfo {author} {\bibfnamefont {S.}~\bibnamefont
  {Ravichandran}}, \bibinfo {author} {\bibfnamefont {E.}~\bibnamefont
  {Meiburg}}, \ and\ \bibinfo {author} {\bibfnamefont {R.}~\bibnamefont
  {Govindarajan}},\ }\href {\doibase 10.1017/jfm.2020.439} {\bibfield
  {journal} {\bibinfo  {journal} {J. Fluid Mech.}\ }\textbf {\bibinfo {volume}
  {899}},\ \bibinfo {pages} {A27} (\bibinfo {year} {2020})}\BibitemShut
  {NoStop}%
\bibitem [{\citenamefont {Shultz}\ \emph {et~al.}(2006)\citenamefont {Shultz},
  \citenamefont {Kanak}, \citenamefont {Straka}, \citenamefont {Trapp},
  \citenamefont {Gordon}, \citenamefont {Zrni{\'c}}, \citenamefont {Bryan},
  \citenamefont {Durant}, \citenamefont {Garrett}, \citenamefont {Klein},\ and\
  \citenamefont {Lilly}}]{Shultz2006}%
  \BibitemOpen
  \bibfield  {author} {\bibinfo {author} {\bibfnamefont {D.~M.}\ \bibnamefont
  {Shultz}}, \bibinfo {author} {\bibfnamefont {K.~M.}\ \bibnamefont {Kanak}},
  \bibinfo {author} {\bibfnamefont {J.~M.}\ \bibnamefont {Straka}}, \bibinfo
  {author} {\bibfnamefont {R.~J.}\ \bibnamefont {Trapp}}, \bibinfo {author}
  {\bibfnamefont {B.~A.}\ \bibnamefont {Gordon}}, \bibinfo {author}
  {\bibfnamefont {D.~S.}\ \bibnamefont {Zrni{\'c}}}, \bibinfo {author}
  {\bibfnamefont {G.~H.}\ \bibnamefont {Bryan}}, \bibinfo {author}
  {\bibfnamefont {A.~J.}\ \bibnamefont {Durant}}, \bibinfo {author}
  {\bibfnamefont {T.~J.}\ \bibnamefont {Garrett}}, \bibinfo {author}
  {\bibfnamefont {P.~M.}\ \bibnamefont {Klein}}, \ and\ \bibinfo {author}
  {\bibfnamefont {D.~K.}\ \bibnamefont {Lilly}},\ }\href {\doibase
  10.1175/JAS3758.1} {\bibfield  {journal} {\bibinfo  {journal} {Journal of the
  Atmospheric Sciences}\ }\textbf {\bibinfo {volume} {63}},\ \bibinfo {pages}
  {2409} (\bibinfo {year} {2006})}\BibitemShut {NoStop}%
\bibitem [{\citenamefont {Kanak}\ and\ \citenamefont
  {Straka}(2006)}]{Kanak2006}%
  \BibitemOpen
  \bibfield  {author} {\bibinfo {author} {\bibfnamefont {K.~M.}\ \bibnamefont
  {Kanak}}\ and\ \bibinfo {author} {\bibfnamefont {J.~M.}\ \bibnamefont
  {Straka}},\ }\href {\doibase 10.1002/asl.121} {\bibfield  {journal} {\bibinfo
   {journal} {Atmospheric Science Letters}\ }\textbf {\bibinfo {volume} {7}},\
  \bibinfo {pages} {2} (\bibinfo {year} {2006})}\BibitemShut {NoStop}%
\bibitem [{\citenamefont {Kanak}\ \emph {et~al.}(2008)\citenamefont {Kanak},
  \citenamefont {Straka},\ and\ \citenamefont {Schultz}}]{Kanak2008}%
  \BibitemOpen
  \bibfield  {author} {\bibinfo {author} {\bibfnamefont {K.~M.}\ \bibnamefont
  {Kanak}}, \bibinfo {author} {\bibfnamefont {J.~M.}\ \bibnamefont {Straka}}, \
  and\ \bibinfo {author} {\bibfnamefont {D.~M.}\ \bibnamefont {Schultz}},\
  }\href {\doibase 10.1175/2007JAS2469.1} {\bibfield  {journal} {\bibinfo
  {journal} {Journal of the Atmospheric Sciences}\ }\textbf {\bibinfo {volume}
  {65}},\ \bibinfo {pages} {1606} (\bibinfo {year} {2008})}\BibitemShut
  {NoStop}%
\bibitem [{\citenamefont {Garrett}\ \emph {et~al.}(2010)\citenamefont
  {Garrett}, \citenamefont {Schmidt}, \citenamefont {Kihlgren},\ and\
  \citenamefont {Cornet}}]{Garrett2010}%
  \BibitemOpen
  \bibfield  {author} {\bibinfo {author} {\bibfnamefont {T.~J.}\ \bibnamefont
  {Garrett}}, \bibinfo {author} {\bibfnamefont {C.~T.}\ \bibnamefont
  {Schmidt}}, \bibinfo {author} {\bibfnamefont {S.}~\bibnamefont {Kihlgren}}, \
  and\ \bibinfo {author} {\bibfnamefont {C.}~\bibnamefont {Cornet}},\ }\href
  {\doibase 10.1175/2010JAS3513.1} {\bibfield  {journal} {\bibinfo  {journal}
  {Journal of the Atmospheric Sciences}\ }\textbf {\bibinfo {volume} {67}},\
  \bibinfo {pages} {3891} (\bibinfo {year} {2010})}\BibitemShut {NoStop}%
\bibitem [{Note1()}]{Note1}%
  \BibitemOpen
  \bibinfo {note}
  {Https://cloudatlas.wmo.int/en/clouds-supplementary-features-asperitas.html}\BibitemShut
  {NoStop}%
\bibitem [{Ave(2009)}]{AveMaria}%
  \BibitemOpen
  \href {{ {{https://commons.wikimedia.org/w/index.php?curid=6581290}}}}
  {\bibfield  {journal} {\bibinfo  {journal} {Wikimedia Commons}\ } (\bibinfo
  {year} {2009})}\BibitemShut {NoStop}%
\bibitem [{Aga(2008)}]{Agathaman}%
  \BibitemOpen
  \href
  {{https://commons.wikimedia.org/wiki/File:Stratocumulus_stratiformis_opacus_lacunosus_undulatus_asperitas.jpg}}
  {\bibfield  {journal} {\bibinfo  {journal} {Wikimedia Commons}\ } (\bibinfo
  {year} {2008})}\BibitemShut {NoStop}%
\bibitem [{\citenamefont {Harrison}\ \emph {et~al.}(2017)\citenamefont
  {Harrison}, \citenamefont {Pretor-Pinney}, \citenamefont {Marlton},
  \citenamefont {Anderson}, \citenamefont {Kirshbaum},\ and\ \citenamefont
  {Hogan}}]{GilesHarrison2017}%
  \BibitemOpen
  \bibfield  {author} {\bibinfo {author} {\bibfnamefont {R.~G.}\ \bibnamefont
  {Harrison}}, \bibinfo {author} {\bibfnamefont {G.}~\bibnamefont
  {Pretor-Pinney}}, \bibinfo {author} {\bibfnamefont {G.~J.}\ \bibnamefont
  {Marlton}}, \bibinfo {author} {\bibfnamefont {G.~D.}\ \bibnamefont
  {Anderson}}, \bibinfo {author} {\bibfnamefont {D.~J.}\ \bibnamefont
  {Kirshbaum}}, \ and\ \bibinfo {author} {\bibfnamefont {R.~J.}\ \bibnamefont
  {Hogan}},\ }\href {\doibase 10.1002/wea.2996} {\bibfield  {journal} {\bibinfo
   {journal} {Weather}\ }\textbf {\bibinfo {volume} {72}},\ \bibinfo {pages}
  {132} (\bibinfo {year} {2017})}\BibitemShut {NoStop}%
\bibitem [{\citenamefont {Burns}\ and\ \citenamefont
  {Meiburg}(2012)}]{Burns2012}%
  \BibitemOpen
  \bibfield  {author} {\bibinfo {author} {\bibfnamefont {P.}~\bibnamefont
  {Burns}}\ and\ \bibinfo {author} {\bibfnamefont {E.}~\bibnamefont
  {Meiburg}},\ }\href {\doibase 10.1017/jfm.2011.474} {\bibfield  {journal}
  {\bibinfo  {journal} {Journal of Fluid Mechanics}\ }\textbf {\bibinfo
  {volume} {691}},\ \bibinfo {pages} {279} (\bibinfo {year}
  {2012})}\BibitemShut {NoStop}%
\bibitem [{\citenamefont {Burns}\ and\ \citenamefont
  {Meiburg}(2014)}]{Burns2014}%
  \BibitemOpen
  \bibfield  {author} {\bibinfo {author} {\bibfnamefont {P.}~\bibnamefont
  {Burns}}\ and\ \bibinfo {author} {\bibfnamefont {E.}~\bibnamefont
  {Meiburg}},\ }\href {\doibase 10.1017/jfm.2014.645} {\bibfield  {journal}
  {\bibinfo  {journal} {Journal of Fluid Mechanics}\ }\textbf {\bibinfo
  {volume} {762}},\ \bibinfo {pages} {156} (\bibinfo {year}
  {2014})}\BibitemShut {NoStop}%
\bibitem [{\citenamefont {Yu}\ \emph {et~al.}(2013)\citenamefont {Yu},
  \citenamefont {Hsu},\ and\ \citenamefont {Balachandar}}]{Yu2013}%
  \BibitemOpen
  \bibfield  {author} {\bibinfo {author} {\bibfnamefont {X.}~\bibnamefont
  {Yu}}, \bibinfo {author} {\bibfnamefont {T.-J.}\ \bibnamefont {Hsu}}, \ and\
  \bibinfo {author} {\bibfnamefont {S.}~\bibnamefont {Balachandar}},\ }\href
  {\doibase 10.1029/2012JC008255} {\bibfield  {journal} {\bibinfo  {journal}
  {Journal of Geophysical Research: Oceans}\ }\textbf {\bibinfo {volume}
  {118}},\ \bibinfo {pages} {256} (\bibinfo {year} {2013})}\BibitemShut
  {NoStop}%
\bibitem [{\citenamefont {Yu}\ \emph {et~al.}(2014)\citenamefont {Yu},
  \citenamefont {Hsu},\ and\ \citenamefont {Balachandar}}]{Yu2014}%
  \BibitemOpen
  \bibfield  {author} {\bibinfo {author} {\bibfnamefont {X.}~\bibnamefont
  {Yu}}, \bibinfo {author} {\bibfnamefont {T.-J.}\ \bibnamefont {Hsu}}, \ and\
  \bibinfo {author} {\bibfnamefont {S.}~\bibnamefont {Balachandar}},\ }\href
  {\doibase 10.1002/2014JC010123} {\bibfield  {journal} {\bibinfo  {journal}
  {Journal of Geophysical Research: Oceans}\ }\textbf {\bibinfo {volume}
  {119}},\ \bibinfo {pages} {8141} (\bibinfo {year} {2014})}\BibitemShut
  {NoStop}%
\bibitem [{\citenamefont {Carazzo}\ and\ \citenamefont
  {Jellinek}(2012)}]{Carazzo2012}%
  \BibitemOpen
  \bibfield  {author} {\bibinfo {author} {\bibfnamefont {G.}~\bibnamefont
  {Carazzo}}\ and\ \bibinfo {author} {\bibfnamefont {M.~A.}\ \bibnamefont
  {Jellinek}},\ }\href {\doibase 10.1016/j.epsl.2012.01.025} {\bibfield
  {journal} {\bibinfo  {journal} {Earth and Planetary Science Letters}\
  }\textbf {\bibinfo {volume} {325-326}},\ \bibinfo {pages} {39} (\bibinfo
  {year} {2012})}\BibitemShut {NoStop}%
\bibitem [{\citenamefont {Carazzo}\ and\ \citenamefont
  {Jellinek}(2013)}]{Carazzo2013}%
  \BibitemOpen
  \bibfield  {author} {\bibinfo {author} {\bibfnamefont {G.}~\bibnamefont
  {Carazzo}}\ and\ \bibinfo {author} {\bibfnamefont {A.~M.}\ \bibnamefont
  {Jellinek}},\ }\href {\doibase 10.1002/jgrb.50155} {\bibfield  {journal}
  {\bibinfo  {journal} {Journal of Geophysical Research: Solid Earth}\ }\textbf
  {\bibinfo {volume} {118}},\ \bibinfo {pages} {1420} (\bibinfo {year}
  {2013})}\BibitemShut {NoStop}%
\bibitem [{\citenamefont {Radko}\ \emph {et~al.}(2015)\citenamefont {Radko},
  \citenamefont {Ball}, \citenamefont {Colosi},\ and\ \citenamefont
  {Flanagan}}]{Radko2015}%
  \BibitemOpen
  \bibfield  {author} {\bibinfo {author} {\bibfnamefont {T.}~\bibnamefont
  {Radko}}, \bibinfo {author} {\bibfnamefont {J.}~\bibnamefont {Ball}},
  \bibinfo {author} {\bibfnamefont {J.}~\bibnamefont {Colosi}}, \ and\ \bibinfo
  {author} {\bibfnamefont {J.}~\bibnamefont {Flanagan}},\ }\href {\doibase
  10.1175/JPO-D-15-0051.1} {\bibfield  {journal} {\bibinfo  {journal} {Journal
  of Physical Oceanography}\ }\textbf {\bibinfo {volume} {45}},\ \bibinfo
  {pages} {3155} (\bibinfo {year} {2015})}\BibitemShut {NoStop}%
\bibitem [{\citenamefont {Konopliv}\ \emph {et~al.}(2018)\citenamefont
  {Konopliv}, \citenamefont {Lesshafft},\ and\ \citenamefont
  {Meiburg}}]{Konopliv2018}%
  \BibitemOpen
  \bibfield  {author} {\bibinfo {author} {\bibfnamefont {N.}~\bibnamefont
  {Konopliv}}, \bibinfo {author} {\bibfnamefont {L.}~\bibnamefont {Lesshafft}},
  \ and\ \bibinfo {author} {\bibfnamefont {E.}~\bibnamefont {Meiburg}},\ }\href
  {\doibase 10.1017/jfm.2018.432} {\bibfield  {journal} {\bibinfo  {journal}
  {Journal of Fluid Mechanics}\ }\textbf {\bibinfo {volume} {849}},\ \bibinfo
  {pages} {902} (\bibinfo {year} {2018})}\BibitemShut {NoStop}%
\bibitem [{\citenamefont {Garaud}\ \emph {et~al.}(2019)\citenamefont {Garaud},
  \citenamefont {Kumar},\ and\ \citenamefont {Sridhar}}]{Garaud2019}%
  \BibitemOpen
  \bibfield  {author} {\bibinfo {author} {\bibfnamefont {P.}~\bibnamefont
  {Garaud}}, \bibinfo {author} {\bibfnamefont {A.}~\bibnamefont {Kumar}}, \
  and\ \bibinfo {author} {\bibfnamefont {J.}~\bibnamefont {Sridhar}},\ }\href
  {\doibase 10.3847/1538-4357/ab232f} {\bibfield  {journal} {\bibinfo
  {journal} {The Astrophysical Journal}\ }\textbf {\bibinfo {volume} {879}},\
  \bibinfo {pages} {60} (\bibinfo {year} {2019})}\BibitemShut {NoStop}%
\bibitem [{\citenamefont {Sichani}\ \emph {et~al.}(2020)\citenamefont
  {Sichani}, \citenamefont {Marchioli}, \citenamefont {Zonta},\ and\
  \citenamefont {Soldati}}]{Sichani2020}%
  \BibitemOpen
  \bibfield  {author} {\bibinfo {author} {\bibfnamefont {P.~H.}\ \bibnamefont
  {Sichani}}, \bibinfo {author} {\bibfnamefont {C.}~\bibnamefont {Marchioli}},
  \bibinfo {author} {\bibfnamefont {F.}~\bibnamefont {Zonta}}, \ and\ \bibinfo
  {author} {\bibfnamefont {A.}~\bibnamefont {Soldati}},\ }\href {\doibase
  10.1115/1.4048342} {\bibfield  {journal} {\bibinfo  {journal} {Journal of
  Fluids Engineering}\ }\textbf {\bibinfo {volume} {142}} (\bibinfo {year}
  {2020}),\ 10.1115/1.4048342}\BibitemShut {NoStop}%
\bibitem [{asp(2011)}]{asperitas_conditions}%
  \BibitemOpen
  \href {https://cloudappreciationsociety.org/asperatus-update} {\bibfield
  {journal} {\bibinfo  {journal} {{Cloud Appreciation Society}}\ } (\bibinfo
  {year} {2011})}\BibitemShut {NoStop}%
\bibitem [{\citenamefont {Maxey}\ and\ \citenamefont
  {Riley}(1983)}]{Maxey1983}%
  \BibitemOpen
  \bibfield  {author} {\bibinfo {author} {\bibfnamefont {M.~R.}\ \bibnamefont
  {Maxey}}\ and\ \bibinfo {author} {\bibfnamefont {J.~J.}\ \bibnamefont
  {Riley}},\ }\href {\doibase 10.1063/1.864230} {\bibfield  {journal} {\bibinfo
   {journal} {Physics of Fluids}\ }\textbf {\bibinfo {volume} {26}},\ \bibinfo
  {pages} {883} (\bibinfo {year} {1983})}\BibitemShut {NoStop}%
\bibitem [{\citenamefont {Pruppacher}\ and\ \citenamefont
  {Klett}(2010)}]{Pruppacher2010}%
  \BibitemOpen
  \bibfield  {author} {\bibinfo {author} {\bibfnamefont {H.}~\bibnamefont
  {Pruppacher}}\ and\ \bibinfo {author} {\bibfnamefont {J.}~\bibnamefont
  {Klett}},\ }\href {\doibase 10.1007/978-0-306-48100-0_2} {\emph {\bibinfo
  {title} {Microstructure of Atmospheric Clouds and Precipitation}}}\ (\bibinfo
   {publisher} {Springer Netherlands},\ \bibinfo {year} {2010})\ pp.\ \bibinfo
  {pages} {10--73}\BibitemShut {NoStop}%
\bibitem [{\citenamefont {Hernandez-Duenas}\ \emph {et~al.}(2013)\citenamefont
  {Hernandez-Duenas}, \citenamefont {Majda}, \citenamefont {Smith},\ and\
  \citenamefont {Stechmann}}]{Hernandez2013}%
  \BibitemOpen
  \bibfield  {author} {\bibinfo {author} {\bibfnamefont {G.}~\bibnamefont
  {Hernandez-Duenas}}, \bibinfo {author} {\bibfnamefont {A.~J.}\ \bibnamefont
  {Majda}}, \bibinfo {author} {\bibfnamefont {L.~M.}\ \bibnamefont {Smith}}, \
  and\ \bibinfo {author} {\bibfnamefont {S.~N.}\ \bibnamefont {Stechmann}},\
  }\href {\doibase 10.1017/jfm.2012.597} {\bibfield  {journal} {\bibinfo
  {journal} {J. Fluid Mech.}\ }\textbf {\bibinfo {volume} {717}},\ \bibinfo
  {pages} {576} (\bibinfo {year} {2013})}\BibitemShut {NoStop}%
\bibitem [{\citenamefont {Pauluis}\ and\ \citenamefont
  {Schumacher}(2010)}]{Pauluis2010}%
  \BibitemOpen
  \bibfield  {author} {\bibinfo {author} {\bibfnamefont {O.}~\bibnamefont
  {Pauluis}}\ and\ \bibinfo {author} {\bibfnamefont {J.}~\bibnamefont
  {Schumacher}},\ }\href {http://projecteuclid.org/euclid.cms/1266935024}
  {\bibfield  {journal} {\bibinfo  {journal} {Communications in Mathematical
  Sciences}\ }\textbf {\bibinfo {volume} {8}},\ \bibinfo {pages} {295}
  (\bibinfo {year} {2010})}\BibitemShut {NoStop}%
\bibitem [{\citenamefont {Vallis}\ \emph {et~al.}(2019)\citenamefont {Vallis},
  \citenamefont {Parker},\ and\ \citenamefont {Tobias}}]{Vallis2019}%
  \BibitemOpen
  \bibfield  {author} {\bibinfo {author} {\bibfnamefont {G.~K.}\ \bibnamefont
  {Vallis}}, \bibinfo {author} {\bibfnamefont {D.~J.}\ \bibnamefont {Parker}},
  \ and\ \bibinfo {author} {\bibfnamefont {S.~M.}\ \bibnamefont {Tobias}},\
  }\href {\doibase 10.1017/jfm.2018.954} {\bibfield  {journal} {\bibinfo
  {journal} {J. Fluid Mech.}\ }\textbf {\bibinfo {volume} {862}},\ \bibinfo
  {pages} {162} (\bibinfo {year} {2019})}\BibitemShut {NoStop}%
\bibitem [{\citenamefont {Devenish}\ \emph {et~al.}(2012)\citenamefont
  {Devenish}, \citenamefont {Bartello}, \citenamefont {Brenguier},
  \citenamefont {Collins}, \citenamefont {Grabowski}, \citenamefont
  {IJzermans}, \citenamefont {Malinowski}, \citenamefont {Reeks}, \citenamefont
  {Vassilicos}, \citenamefont {Wang},\ and\ \citenamefont
  {Warhaft}}]{Devenish2012}%
  \BibitemOpen
  \bibfield  {author} {\bibinfo {author} {\bibfnamefont {B.~J.}\ \bibnamefont
  {Devenish}}, \bibinfo {author} {\bibfnamefont {P.}~\bibnamefont {Bartello}},
  \bibinfo {author} {\bibfnamefont {J.-L.}\ \bibnamefont {Brenguier}}, \bibinfo
  {author} {\bibfnamefont {L.~R.}\ \bibnamefont {Collins}}, \bibinfo {author}
  {\bibfnamefont {W.~W.}\ \bibnamefont {Grabowski}}, \bibinfo {author}
  {\bibfnamefont {R.~H.~A.}\ \bibnamefont {IJzermans}}, \bibinfo {author}
  {\bibfnamefont {S.~P.}\ \bibnamefont {Malinowski}}, \bibinfo {author}
  {\bibfnamefont {M.~W.}\ \bibnamefont {Reeks}}, \bibinfo {author}
  {\bibfnamefont {J.~C.}\ \bibnamefont {Vassilicos}}, \bibinfo {author}
  {\bibfnamefont {L.-P.}\ \bibnamefont {Wang}}, \ and\ \bibinfo {author}
  {\bibfnamefont {Z.}~\bibnamefont {Warhaft}},\ }\href {\doibase
  10.1002/qj.1897} {\bibfield  {journal} {\bibinfo  {journal} {Quarterly
  Journal of the Royal Meteorological Society}\ }\textbf {\bibinfo {volume}
  {138}},\ \bibinfo {pages} {1401} (\bibinfo {year} {2012})}\BibitemShut
  {NoStop}%
\bibitem [{\citenamefont {Carpenter}\ \emph {et~al.}(2011)\citenamefont
  {Carpenter}, \citenamefont {Tedford}, \citenamefont {Heifetz},\ and\
  \citenamefont {Lawrence}}]{Carpenter2011}%
  \BibitemOpen
  \bibfield  {author} {\bibinfo {author} {\bibfnamefont {J.~R.}\ \bibnamefont
  {Carpenter}}, \bibinfo {author} {\bibfnamefont {E.~W.}\ \bibnamefont
  {Tedford}}, \bibinfo {author} {\bibfnamefont {E.}~\bibnamefont {Heifetz}}, \
  and\ \bibinfo {author} {\bibfnamefont {G.~A.}\ \bibnamefont {Lawrence}},\
  }\href {\doibase 10.1115/1.4007909} {\bibfield  {journal} {\bibinfo
  {journal} {Applied Mechanics Reviews}\ }\textbf {\bibinfo {volume} {64}},\
  \bibinfo {pages} {60801} (\bibinfo {year} {2011})}\BibitemShut {NoStop}%
\bibitem [{\citenamefont {Kurganov}\ and\ \citenamefont
  {Tadmor}(2000)}]{Kurganov2000}%
  \BibitemOpen
  \bibfield  {author} {\bibinfo {author} {\bibfnamefont {A.}~\bibnamefont
  {Kurganov}}\ and\ \bibinfo {author} {\bibfnamefont {E.}~\bibnamefont
  {Tadmor}},\ }\href@noop {} {\bibfield  {journal} {\bibinfo  {journal} {J.
  Comput. Phys.}\ }\textbf {\bibinfo {volume} {160}},\ \bibinfo {pages} {241}
  (\bibinfo {year} {2000})}\BibitemShut {NoStop}%
\bibitem [{\citenamefont {Traxler}\ \emph {et~al.}(2011)\citenamefont
  {Traxler}, \citenamefont {Stellmach}, \citenamefont {Garaud}, \citenamefont
  {Radko},\ and\ \citenamefont {Brummell}}]{TRAXLER2011}%
  \BibitemOpen
  \bibfield  {author} {\bibinfo {author} {\bibfnamefont {A.}~\bibnamefont
  {Traxler}}, \bibinfo {author} {\bibfnamefont {S.}~\bibnamefont {Stellmach}},
  \bibinfo {author} {\bibfnamefont {P.}~\bibnamefont {Garaud}}, \bibinfo
  {author} {\bibfnamefont {T.}~\bibnamefont {Radko}}, \ and\ \bibinfo {author}
  {\bibfnamefont {N.}~\bibnamefont {Brummell}},\ }\href {\doibase
  10.1017/jfm.2011.98} {\bibfield  {journal} {\bibinfo  {journal} {Journal of
  Fluid Mechanics}\ }\textbf {\bibinfo {volume} {677}},\ \bibinfo {pages} {530}
  (\bibinfo {year} {2011})}\BibitemShut {NoStop}%
\bibitem [{\citenamefont {Fu}\ \emph {et~al.}(2012)\citenamefont {Fu},
  \citenamefont {Woo},\ and\ \citenamefont {Chen}}]{Fu2012}%
  \BibitemOpen
  \bibfield  {author} {\bibinfo {author} {\bibfnamefont {N.}~\bibnamefont
  {Fu}}, \bibinfo {author} {\bibfnamefont {M.~W.}\ \bibnamefont {Woo}}, \ and\
  \bibinfo {author} {\bibfnamefont {X.~D.}\ \bibnamefont {Chen}},\ }\href
  {\doibase 10.1080/07373937.2012.708002} {\bibfield  {journal} {\bibinfo
  {journal} {Drying Technology}\ }\textbf {\bibinfo {volume} {30}},\ \bibinfo
  {pages} {1771} (\bibinfo {year} {2012})}\BibitemShut {NoStop}%
\bibitem [{\citenamefont {Ji}\ \emph {et~al.}(2010)\citenamefont {Ji},
  \citenamefont {Zeng}, \citenamefont {Ji}, \citenamefont {Yang}, \citenamefont
  {Liu},\ and\ \citenamefont {Li}}]{Ji2010}%
  \BibitemOpen
  \bibfield  {author} {\bibinfo {author} {\bibfnamefont {J.}~\bibnamefont
  {Ji}}, \bibinfo {author} {\bibfnamefont {P.}~\bibnamefont {Zeng}}, \bibinfo
  {author} {\bibfnamefont {S.}~\bibnamefont {Ji}}, \bibinfo {author}
  {\bibfnamefont {W.}~\bibnamefont {Yang}}, \bibinfo {author} {\bibfnamefont
  {H.}~\bibnamefont {Liu}}, \ and\ \bibinfo {author} {\bibfnamefont
  {Y.}~\bibnamefont {Li}},\ }\href {\doibase 10.1016/j.cattod.2010.03.074}
  {\bibfield  {journal} {\bibinfo  {journal} {Catalysis Today}\ }\textbf
  {\bibinfo {volume} {158}},\ \bibinfo {pages} {305} (\bibinfo {year}
  {2010})}\BibitemShut {NoStop}%
\bibitem [{\citenamefont {Hu}\ \emph {et~al.}(2013)\citenamefont {Hu},
  \citenamefont {Liu}, \citenamefont {Wang}, \citenamefont {Liu}, \citenamefont
  {Liu}, \citenamefont {Jing}, \citenamefont {Yu}, \citenamefont {Liu},\ and\
  \citenamefont {Zhang}}]{Hu2013}%
  \BibitemOpen
  \bibfield  {author} {\bibinfo {author} {\bibfnamefont {W.}~\bibnamefont
  {Hu}}, \bibinfo {author} {\bibfnamefont {B.}~\bibnamefont {Liu}}, \bibinfo
  {author} {\bibfnamefont {Q.}~\bibnamefont {Wang}}, \bibinfo {author}
  {\bibfnamefont {Y.}~\bibnamefont {Liu}}, \bibinfo {author} {\bibfnamefont
  {Y.}~\bibnamefont {Liu}}, \bibinfo {author} {\bibfnamefont {P.}~\bibnamefont
  {Jing}}, \bibinfo {author} {\bibfnamefont {S.}~\bibnamefont {Yu}}, \bibinfo
  {author} {\bibfnamefont {L.}~\bibnamefont {Liu}}, \ and\ \bibinfo {author}
  {\bibfnamefont {J.}~\bibnamefont {Zhang}},\ }\href {\doibase
  10.1039/c3cc42687d} {\bibfield  {journal} {\bibinfo  {journal} {Chemical
  Communications}\ }\textbf {\bibinfo {volume} {49}},\ \bibinfo {pages} {7596}
  (\bibinfo {year} {2013})}\BibitemShut {NoStop}%
\bibitem [{\citenamefont {Radko}(2012)}]{Radko2012}%
  \BibitemOpen
  \bibfield  {author} {\bibinfo {author} {\bibfnamefont {T.}~\bibnamefont
  {Radko}},\ }\href {\doibase 10.1017/CBO9781139034173} {\bibfield  {journal}
  {\bibinfo  {journal} {Double-Diffusive Convection}\ }\textbf {\bibinfo
  {volume} {9780521880}},\ \bibinfo {pages} {1} (\bibinfo {year}
  {2012})}\BibitemShut {NoStop}%
\bibitem [{\citenamefont {Ouillon}\ \emph {et~al.}(2020)\citenamefont
  {Ouillon}, \citenamefont {Edel}, \citenamefont {Garaud},\ and\ \citenamefont
  {Meiburg}}]{Ouillon2020}%
  \BibitemOpen
  \bibfield  {author} {\bibinfo {author} {\bibfnamefont {R.}~\bibnamefont
  {Ouillon}}, \bibinfo {author} {\bibfnamefont {P.}~\bibnamefont {Edel}},
  \bibinfo {author} {\bibfnamefont {P.}~\bibnamefont {Garaud}}, \ and\ \bibinfo
  {author} {\bibfnamefont {E.}~\bibnamefont {Meiburg}},\ }\href {\doibase
  10.1017/jfm.2020.527} {\bibfield  {journal} {\bibinfo  {journal} {Journal of
  Fluid Mechanics}\ }\textbf {\bibinfo {volume} {901}} (\bibinfo {year}
  {2020}),\ 10.1017/jfm.2020.527}\BibitemShut {NoStop}%
\end{thebibliography}
%

\begin{acknowledgments}
We dedicate this work to Prof. Roddam Narasimha, who
died in December 2020 after a career in science spanning more than
six decades and three generations of students and colleagues including
both present authors. Computations were performed on the ICTS clusters Mowgli and Contra. RG acknowledges support of the Department of Atomic
Energy, Government of India, under project no. RTI4001. SR gratefully
acknowledges support through the Swedish Research Council grant no.
638-2013-9243. 
\end{acknowledgments}

\end{document}